\documentclass[aps, prd, showkeys, singlecolumn, nofootinbib, floatfix]{revtex4-2}

\usepackage{float}
\usepackage{amssymb}
\usepackage{amsmath}
\usepackage{graphicx}
\usepackage{dcolumn}
\usepackage{bm}
\usepackage[frak=esstix]{mathalpha}
\usepackage{soul, xpatch, caption, subcaption}
\usepackage[normalem]{ulem}
\usepackage[dvipsnames]{xcolor}
\usepackage[colorlinks=true,allcolors = magenta]{hyperref}
\usepackage[multiple]{footmisc}
\usepackage{tensor}
\usepackage[most]{tcolorbox}

\graphicspath{{figures/}}

\begin{document}

\title{Causality and stability of magnetohydrodynamics for an ultrarelativistic locally neutral two-component gas}

\author{Caio V.~P.~de Brito}
\email{caio\_brito@id.uff.br}

\author{Khwahish Kushwah}
\email{khwahish\_kushwah@id.uff.br}

\author{Gabriel S.~Denicol}
\email{gsdenicol@id.uff.br}

\begin{abstract}
We investigate the causality and stability of the relativistic theory of magnetohydrodynamics derived in Phys.~Rev.~D 109, 096021 (2024) to describe a locally neutral two-component plasma of massless particles. 
We show that this formalism is linearly causal and stable around global equilibrium, for any value of the magnetic field and discuss its qualitative differences to the traditional Israel-Stewart formalism in the linear regime. Finally, we compare this framework with the magnetohydrodynamic model used in the study of astrophysical plasmas, in which only the longitudinal component of the shear-stress tensor is considered. We discuss the domain of applicability of this type of framework in the context of ultrarelativistic heavy-ion collisions.
\end{abstract}

\affiliation{Instituto de F\'{\i}sica, Universidade Federal Fluminense \\
Av.~Gal.~Milton Tavares de Souza, S/N, 24210-346, Gragoatá, Niter\'{o}i,
Rio de Janeiro, Brazil}

\maketitle

\section{Introduction}

Strong magnetic fields are present in a wide variety of physical systems and play a crucial role in environments where relativistic hydrodynamics is relevant. These fields span from the Earth's magnetic field ($\sim$ 0.5~Gauss) to the extreme fields observed in neutron stars, which can reach magnitudes of $10^{15}$~Gauss \cite{Adhikari:2024bfa, Kiuchi:2015sga, Duncan:1992hi, Price:2006fi}, and non-central ultrarelativistic heavy-ion collisions where fields of the order $10^{18}$--$10^{19}$ Gauss are expected to be produced due to the motion of spectator ions \cite{Skokov:2009qp, Voronyuk:2011jd, Deng:2012pc, Bzdak:2011yy, Bloczynski:2012en, Tuchin:2013apa, Yan:2021zjc, Wang:2021oqq}. In the case of heavy-ion collisions, such strong magnetic fields may significantly influence the early-stage dynamics of the quark-gluon plasma \cite{Rischke:2003mt} produced in these collisions and induce novel transport phenomena with potentially observable consequences in experimental measurements \cite{Miransky:2015ava, Huang:2015oca, Hattori:2016emy, Kharzeev:2015znc, Kharzeev:2020jxw, STAR:2023jdd, Dubla:2020bdz}.

The dynamics of relativistic plasmas in the presence of electromagnetic fields is described by relativistic magnetohydrodynamics, a theory describing the long-wavelength and low-frequency dynamics of matter coupled to electromagnetic fields. This framework is of broad interest, as it is crucial to analyze the evolution of magnetic fields in heavy-ion collisions \cite{Hattori_2022, Tuchin:2013ie, Li:2016tel, Pu:2016ayh, Zheng:2025nra, Roy:2015kma, Hongo:2013cqa, Inghirami:2016iru, Inghirami:2019mkc}, which can have an effect on the transport properties of the system \cite{Hattori:2016lqx, Hattori:2016cnt, Hattori:2016idp, Hattori:2017qih, Li:2017tgi} as well as impact the chiral magnetic effect and other anomalous transport phenomena \cite{Kharzeev:2007jp,Gursoy:2014aka, Son:2009tf, Ammon:2020rvg, Hattori_2022, Fukushima:2008xe, Siddique:2019gqh, Wang:2020qpx}. It is also successful in describing the large-scale structure of the universe~\cite{Armas:2022wvb, Brandenburg:1996fc, refId0}, and the plasma behavior in extreme astrophysical environments such as black hole accretion disks and neutron star mergers \cite{Anile_1990, Armas:2022wvb, Prieto:2015efa, Abramowicz:2011xu}.

Developing a causal and linearly stable (around global equilibrium) theory of relativistic magnetohydrodynamics is a challenge. The main reason is that one needs to derive a second-order version of magnetohydrodynamics, which contains the transient dynamics of the dissipative currents. This endeavor has been pursued by several authors \cite{Denicol:2018rbw, Denicol:2019iyh, Dey:2019vkn, Chen:2019usj, Dash:2020vxk, Singh:2020faa, Panda:2020zhr, Panda:2022imm, Ghosh:2022xtv,Critelli:2014kra, Finazzo:2016mhm, Li:2018ufq, Fukushima:2021got}. One way to do this is to extend the Israel-Stewart derivation of hydrodynamics from the Boltzmann equation, via the method of moments~\cite{israel1979annals,Denicol_Rischke,cercignani2002}, to the Boltzmann-Vlasov equation, as first performed in Refs.~\cite{Denicol:2018rbw,Denicol:2019iyh} for a single component gas.  In Refs.~\cite{Kushwah:2024zgd,Kushwah:2024csd} this procedure was further developed and a theory of non-resistive relativistic magnetohydrodynamics was derived for a \textit{locally neutral} two-component gas made of classical massless particles. In this case, it was shown that the shear-stress tensor does not necessarily satisfy a traditional Israel-Stewart-type equation of motion. Instead, it was found that different components of the shear-stress tensor, decomposed with respect to the direction of the magnetic field, satisfy distinct evolution equations, with transport coefficients that display a significant dependence on the magnetic field. It was also demonstrated that, when the magnetic field becomes large, the shear-stress tensor exhibits oscillatory dynamics that can never be captured by the conventional Israel-Stewart equations. Nevertheless, the causality and linear stability (around global equilibrium) of this new theory of second-order magnetohydrodynamics remains to be verified \footnote{We note that linear stability and causality analyses of magnetohydrodynamics, using the traditional Israel-Stewart theory were performed in Refs.~\cite{Biswas:2020rps, Biswas:2022gwa}. Linear stability analysis of resistive ideal magnetohydrodynamics were performed in Ref.~\cite{Dionysopoulou:2012zv}}. 

In this paper we perform this task and implement a linear stability and causality analysis of the novel magnetohydrodynamic equations derived in Ref.~\cite{Kushwah:2024zgd}. We linearize the equations of motion and decompose them in Fourier space using a complete orthonormal basis. We obtain the dispersion relations and determine the hydrodynamic and nonhydrodynamic modes of the theory for perturbations that are either parallel or transverse to the magnetic field. We show that this formulation of relativistic magnetohydrodynamics is linearly causal and stable, and determined how the hydrodynamic modes depend on the magnetic field, temperature, and cross-sections. We also demonstrate that the hydrodynamic modes that couple to the perturbations of the shear-stress tensor that are longitudinal with respect to the magnetic field do not display any dependence on the magnetic field. On the other hand, the hydrodynamic modes that couple to the transverse perturbations of the shear-stress tensor display a strong dependence on the magnetic field -- becoming almost non-dissipative in the limit of large magnetic fields. We also analyze the modes of a simplified limit of non-resistive magnetohydrodynamics that is expected to be applicable in the limit of strong magnetic fields \cite{Chandra:2015iza}. In this regime, one assumes that the shear-stress tensor is dominantly expressed in terms of its components parallel to the magnetic field. We discuss the domain of applicability of this approximation in the linear regime. 

This paper is organized into the following sections. In Sec.~\ref{sec:fundamentals-mhd}, we discuss the basic equations of magnetohydrodynamics. In Sec.~\ref{sec:linear-mhd}, we linearize the conservation laws as well as the equations of motion for the shear-stress tensor and transform them into Fourier space. These are then decomposed into an orthonormal basis of 4-vectors and rescaled into dimensionless quantities in Sec.~\ref{subsec:resc-linear-mhd-four}. Equipped with these equations, we perform a causality and stability analysis in Sec.~\ref{sec:stab-analysis} for longitudinal and transverse perturbations with respect to the magnetic field. These modes are then compared to the limits of the magnetohydrodynamic framework proposed in~\cite{Chandra:2015iza} in Sec.~\ref{sec:long-limit-mhd}. Finally, in Sec.~\ref{sec:conclusions}, we summarize our main findings.

Throughout this work, we adopt natural units, i.e., $\hbar = c = k_B = 1$,  and the background spacetime is considered to be flat Minkowski space, characterized by the metric tensor $g_{\mu\nu} = \mathrm{diag}(+1, -1, -1, -1)$.

\section{Fundamentals of relativistic magnetohydrodynamics}
\label{sec:fundamentals-mhd}
In this work we consider a locally neutral, non-resistive, ultra-relativistic plasma. The fundamental relations that govern this system are the continuity equations that describe the conservation of energy and momentum, 
\begin{equation}
\label{eq:cons-energ-mom}
\partial_\mu T^{\mu \nu}=0, 
\end{equation}
where $T^{\mu\nu}$ is the energy-momentum tensor. In the context of magnetohydrodynamics, the \textit{total} energy-momentum tensor of the system carries an electromagnetic contribution besides the usual fluid-dynamical term, 
\begin{equation}
T^{\mu\nu} = T^{\mu\nu}_{\mathrm{EM}} + T^{\mu\nu}_{\mathrm{fluid}},
\end{equation}
where $T^{\mu\nu}_{\mathrm{EM}}$ and $T^{\mu\nu}_{\mathrm{fluid}}$ denote the electromagnetic and fluid contributions, respectively. In general, the latter is given by \cite{landau1959} 
\begin{equation}
T^{\mu\nu}_{\mathrm{fluid}} = \varepsilon u^\mu u^\nu-\Delta^{\mu\nu}P  + \pi^{\mu\nu},
\label{eq:tmunu-fluid}
\end{equation}
where $\varepsilon$ is the energy density, $u^\mu$ is the fluid 4-velocity, a normalized time-like 4-vector, $u_\mu u^\mu = 1$, $P$ is the thermodynamic pressure, and $\pi^{\mu\nu}$ is the shear-stress tensor.  We also introduced the projection operator $\Delta^{\mu\nu} = g^{\mu \nu} - u^\mu u^\nu$. Above, we fixed the velocity field using Landau matching conditions \cite{landau1959}, in which $u^\mu$ is determined as the time-like eigenvector of the energy-momentum tensor, i.e., $T^{\mu \nu }u_{\nu }=\varepsilon u^\mu$.  Since we consider an ultra-relativistic fluid, we have neglected the bulk viscous pressure contribution to the energy-momentum tensor.

The electromagnetic contribution to the energy-momentum tensor can be expressed as \cite{cercignani2002}, 
\begin{equation}
T^{\mu\nu}_{\mathrm{EM}} = - F^{\mu\lambda}F^{\nu}{}_{\lambda} + \frac{1}{4}
g^{\mu\nu} F^{\alpha\beta} F_{\alpha\beta},
\end{equation}
where $F_{\mu\nu} = \partial_\mu A_\nu - \partial_\nu A_\mu$ is the Faraday tensor, with $A_\mu$ being the electromagnetic 4-potential. The Faraday tensor can be generally decomposed with respect to the fluid 4-velocity in the following way \cite{Ellis:1973jva, gedalin1993},
\begin{equation}
\label{eq:farad-tens}
F^{\mu\nu} = E^\mu u^\nu - E^\nu u^\mu + \epsilon^{\mu\nu\alpha\beta}
u_\alpha B_\beta,
\end{equation}
with $\epsilon^{\mu\nu\alpha\beta}$ being the 4-dimension Levi-Civita symbol. Above, we introduced  the electric and magnetic field 4-vectors, defined as $E^\mu = F^{\mu\nu}u_\nu$ and $B^\mu = \frac{1}{2} \epsilon^{\mu\nu\alpha\beta} F_{\alpha\beta} u_\nu$, respectively. Similarly, the Hodge dual of the Faraday tensor can be tensor decomposed as\footnote{The Hodge dual of the Faraday tensor is usually denoted by $\tilde{F}^{\mu \nu }$ \cite{jackson2012classical}. However, in order to avoid confusion with the Fourier transform in the following sections, in this work we adopt the notation $\mathring{F}^{\mu \nu }$.},  \begin{equation}
\label{eq:farad-hodge}
\mathring{F}^{\mu \nu }\equiv \frac{1}{2}\epsilon ^{\mu \nu \alpha \beta }F_{\alpha \beta }=B^{\mu }u^{\nu }-B^{\nu }u^{\mu }-\epsilon ^{\mu \nu \alpha \beta }u_{\alpha }E_{\beta }.
\end{equation}%
As already mentioned, we consider a non-resistive plasma (infinite electric conductivity) and, thus, assume that the 4-electric field can be neglected $E^\mu \approx 0$. In this case, the electromagnetic contribution to the energy-momentum tensor simplifies to, 
\begin{equation}
\label{eq:EM-tmunu}
T^{\mu\nu}_{\mathrm{EM}} = - B^\mu B^\nu + B^2 u^\mu u^\nu -   \frac{1}{2} B^2g^{\mu\nu},
\end{equation}
where we defined the magnitude (squared) of the magnetic field $B^2 = - B_\mu B^\mu$. Therefore, from Eqs.~\eqref{eq:tmunu-fluid} and \eqref{eq:EM-tmunu}, the \textit{total} energy-momentum becomes,
\begin{equation}
\label{eq:general-mhd-tmunu}
T^{\mu\nu} = \left( \varepsilon + \frac{B^2}{2} \right)
u^\mu u^\nu - \Delta^{\mu\nu} \left( P  + \frac{B^2}{2}
\right) + \pi^{\mu\nu}- B^\mu
B^\nu .
\end{equation}

The evolution of the fluid-dynamical fields is coupled with the magnetic field 4-vector, which is governed by Maxwell's equation for the Hodge dual,
\begin{equation}
\label{eq:hodge-maxwell-eq}
\partial _{\mu }\mathring{F}^{\mu \nu }=0\Longrightarrow u^{\nu }\partial_{\mu }B^{\mu }+B^{\mu }\nabla _{\mu }u^{\nu }-B^{\nu }\theta -D B^\nu  = 0.
\end{equation}%
where $D \equiv u^\mu \partial_\mu$ is the comoving time derivative, $\nabla^\mu \equiv \Delta^{\mu\nu} \partial_\nu$ is the 4-gradient operator and $\theta \equiv \partial_\mu u^\mu$ is the expansion rate. We note that since we restrict our analyses to a plasma with an infinite conductivity, the net-charge 4-current, $J^\mu$, is determined from the Maxwell equation itself, $\partial_\mu F^{\mu\nu}=J^\nu$. This equation was actually already used to write the expression for $T^{\mu \nu}_\mathrm{EM}$ in Eq.~\eqref{eq:EM-tmunu} and will not be considered further in this work.

For the sake of closure, it still is necessary to provide equations for the dissipative currents. In this work, we consider a locally neutral two-component fluid of massless particles with a zero electric field, thoroughly investigated in Ref.~\cite{Kushwah:2024zgd}. In this case, both bulk viscous pressure and diffusion 4-current are identically zero, and dissipative effects stem solely from the shear-stress tensor. Equations of motion for the shear-stress tensor for each particle species (denoted by $\pi^{\mu\nu}_+$ and $\pi^{\mu\nu}_-$ for $+q$ and $-q$ charges, respectively) are individually derived using the Boltzmann-Vlasov equation and further truncated employing the 14-moment approximation \cite{Denicol:2012es, Denicol:2018rbw}. These equations can then be expressed in terms of the \textit{total} and the \textit{relative} shear stress tensor, defined respectively as
\begin{equation}
\pi ^{\mu \nu } \equiv \pi_+^{\mu \nu} + \pi_-^{\mu \nu}, 
\qquad \text{and} \qquad 
\delta \pi^{\mu \nu} \equiv \pi_+^{\mu \nu} - \pi_-^{\mu \nu},
\end{equation}
which satisfy coupled differential equations of motion given by 
\begin{subequations}
\label{eq:coupled-PDEs-shear}
\begin{align}
\Delta _{\alpha \beta }^{\mu \nu }D\pi^{\alpha \beta }+\Sigma \pi ^{\mu
\nu }+\omega_0 b^{\lambda \langle \mu }\delta \pi _{\lambda }^{\nu
\rangle }& =\frac{8}{15}\varepsilon \sigma ^{\mu \nu }-\frac{4}{3}\pi ^{\mu \nu
}\theta -\frac{10}{7}\sigma ^{\lambda \langle \mu }\pi _{\lambda }^{\nu
\rangle }-2\omega ^{\lambda \langle \nu }\pi _{\lambda }^{\mu \rangle },
\label{eq:delta-shear} \\
\Delta _{\alpha \beta }^{\mu \nu }D\delta \pi^{\alpha \beta }+\Sigma' \delta \pi ^{\mu \nu }+\omega_0 b^{\lambda \langle \mu
}\pi _{\lambda }^{\nu \rangle }& =-\frac{4}{3}\delta \pi ^{\mu \nu }\theta -%
\frac{10}{7}\sigma ^{\lambda \langle \mu }\delta \pi _{\lambda }^{\nu
\rangle }-2\omega ^{\lambda \langle \nu }\delta \pi _{\lambda }^{\mu \rangle
},\hspace{1.2cm}
\end{align}
\end{subequations}
where we defined the shear tensor, $\displaystyle{\sigma^{\mu\nu} \equiv \nabla^{\langle\mu} \ u^{\nu\rangle}}$, and the vorticity tensor, $\displaystyle{\omega^{\mu\nu} = (\nabla^{\mu}u^\nu - \nabla^\nu u^\mu)/2}$. We have also employed the notation, $\displaystyle{A^{\langle\mu\nu\rangle} \equiv \Delta^{\mu\nu}_{\alpha\beta}A^{\alpha\beta}}$, with $\displaystyle{\Delta^{\mu\nu}_{\alpha\beta}  \equiv \left(\Delta^{\mu}_\alpha \Delta^{\nu}_\beta + \Delta^{\nu}_\alpha\Delta^{\mu}_\beta - 2/3\Delta^{\mu\nu}\Delta_{\alpha\beta}\right)/2}$.  
Finally, we defined the frequency,
\begin{equation}
\omega_0 = \frac{2 |q| B}{5 T},
\end{equation}
with $T$ being the temperature of the plasma. This frequency is inversely proportional to the Larmor radius of a particle with transverse momentum of order $\sim T$.
Since the total cross-sections corresponding to the interaction between the same particle species are assumed to be identical and constant, i.e., $\sigma_T^{++} = \sigma_T^{--} = \sigma_T$, the coupling between the two species arises solely due to the presence of a magnetic field. Further, the cross-section corresponding to the inter-species interaction is denoted by, $\sigma^{+-}_T = \sigma^{-+}_T $.  Here, $\Sigma$ and $\Sigma^{'}$ are positive definite functions of the cross-sections \cite{Kushwah:2024zgd} and $b^{\mu\nu}$ is an antisymmetric second-rank tensor defined as
\begin{equation}
    b^{\mu\nu} = -\epsilon^{\mu\nu\alpha\beta} u_\alpha b_\beta,
\end{equation}
where $b^\mu = \displaystyle{B^\mu/B}$ is a unitary space-like 4-vector, $b^\mu b_\mu = -1$, in the direction of the magnetic field. By construction, it is also orthogonal to the 4-velocity, $u_\mu b^\mu = 0$.

The primary goal of this work is to determine whether the magneto-fluid-dynamical formulation developed in Ref.~\cite{Kushwah:2024zgd} is well-behaved under \textit{small} perturbations around a global equilibrium state, i.e., if it is stable and causal around global equilibrium. 

\section{Linearized magnetohydrodynamics}
\label{sec:linear-mhd}

We consider \textit{small} deviations of the magneto-fluid-dynamical fields (denoted by $\Delta$) around a global equilibrium state with energy density $\varepsilon_0$, 4-velocity $u^\mu_0$ and zero net-charge,
\begin{eqnarray}
\label{eq:small-pert}
\varepsilon = \varepsilon_0 + \Delta \varepsilon, \ \
u^\mu = u^\mu_0 + \Delta u^\mu, \ \ 
\pi^{\mu\nu} = \Delta \pi^{\mu\nu}, \ \  
\delta \pi^{\mu\nu} = \Delta \delta \pi^{\mu\nu}, \ \  
B^\mu = B_0 b_0^\mu + \Delta B^\mu,
\end{eqnarray}
where one can express $\Delta B^\mu = b^\mu_0 \Delta B + B_0 \Delta b^\mu $. We remark that all perturbations of the shear-stress tensor (and, similarly, the relative shear-stress tensor) can be approximated as orthogonal to the background 4-velocity,
\begin{equation}
\Delta \pi^{\mu\nu} u^0_\mu = \mathcal{O}(\Delta^2) \approx 0.
\end{equation}
The perturbations on the fluid velocity and normalized magnetic field are both orthogonal to their unperturbed counterparts up to first order in perturbations, since both are normalized 4-vectors,
\begin{equation}
u^\mu_0 \Delta u_\mu = \mathcal{O}(\Delta^2) \approx 0, \ 
b^\mu_0 \Delta b_\mu = \mathcal{O}(\Delta^2) \approx 0.
\end{equation}
The background fluid 4-velocity is not orthogonal to perturbations of the magnetic field and vice-versa,
\begin{equation}
u^\mu_0 \Delta b_\mu = - b^\mu_0 \Delta u_\mu + \mathcal{O}(\Delta^2).
\end{equation}
In the following, we linearize the magneto-fluid-dynamical equations, neglecting all terms of second order (or higher) in the perturbations denoted in Eqs.~\eqref{eq:small-pert}.

The continuity equation satisfied by the energy-momentum tensor reduce to,
\begin{eqnarray}
\label{eq:linear-cons-tmunu}
&&\partial _{\mu }\left[ \left( \varepsilon _{0}+P_{0}+B_{0}^{2}\right)
\left( u_{0}^{\mu }\Delta u^{\nu }+\Delta u^{\mu }u_{0}^{\nu
}\right) -B_{0}^{2}\left( b_{0}^{\mu }\Delta b^{\nu }+b_{0}^{\nu
}\Delta b^{\mu }\right) \right. \notag \\
&&\left. -\Delta P\left( \Delta _{0}^{\mu \nu }-3u_{0}^{\mu
}u_{0}^{\nu }\right) -B_{0}\Delta B\left( \Xi _{0}^{\mu \nu
}-u_{0}^{\mu }u_{0}^{\nu }+b_{0}^{\mu }b_{0}^{\nu }\right) +\Delta \pi^{\mu \nu }\right] = \mathcal{O}(\Delta^2) \approx 0,
\end{eqnarray}
where $\Delta^{\mu \nu}_0 = g^{\mu\nu} - u_0^\mu u_0^\nu$ is the projection operator onto the 3-space orthogonal to $u_0^\mu$ and $\Xi _{0}^{\mu \nu}=\Delta^{\mu \nu}_0 + b_0^\mu b_0^\nu$ is the projection operator onto the 2-space orthogonal to $u_0^\mu$ and $b_0^\mu$. The linearized Maxwell's equation for the Hodge dual become,
\begin{align}
\label{eq:maxwell-linear-hodge}
\partial_\mu \Delta \mathring{F}^{\mu\nu}
&= u^\nu_0 \partial_\mu (b^\mu_0 \Delta B + B_0 \Delta b^\mu) + B_0 b^\mu_0 \nabla_\mu^0 \Delta u^\nu - B_0 b^\nu_0 \nabla_\mu^0 \Delta u^\mu - b^\nu_0 D_0\Delta B - B_0 D_0\Delta b^\nu
= \mathcal{O}(\Delta^2) \approx 0.
\end{align}
where we have defined the linearized comoving time derivative $D_0 \equiv u^\mu_0 \partial_\mu$ as well as the linearized spatial derivative $\nabla_0^\mu \equiv \Delta^{\mu \nu}_0 \partial_\nu$.

Finally, the equations of motion for the total and relative shear-stress tensor, Eqs.~\eqref{eq:coupled-PDEs-shear}, reduce to
\begin{subequations}
\label{eq:linearized-shear}
\begin{align}
D_0 \Delta \pi^{\mu \nu } + \Sigma \Delta \pi^{\mu \nu} + \omega_0 b_0^{\lambda \langle \mu } \Delta \delta \pi_\lambda^{\nu \rangle}
&= \frac{4}{15} \varepsilon_0 \left( \nabla^\mu_0 \Delta u^\nu + \nabla^\nu_0 \Delta u^\mu - \frac{2}{3} \Delta^{\mu\nu}_0 \partial_\alpha \Delta u^\alpha \right) + \mathcal{O} (\Delta^2), \label{eq:linear-total-shear} \\
D_0 \Delta \delta \pi^{\mu \nu } + \Sigma \Delta \delta \pi^{\mu \nu} + \omega_0 b_0^{\lambda \langle \mu } \Delta \pi_\lambda^{\nu \rangle}
&= \mathcal{O} (\Delta^2), \label{eq:linear-relative-shear}
\end{align}
\end{subequations}
where $\omega_0 = 2 |q| B_0 / (5T_0)$ now corresponds to the frequency of the unperturbed fluid and we defined $b^{\mu\nu}_0 = - \epsilon^{\mu\nu\alpha\beta} u_\alpha^0 b^0_\beta$.

\subsection{Introducing a new basis in Fourier space}
\label{subsec: Introducing a new basis in Fourier space}

It is practical to express the linearized fluid-dynamical equations in Fourier space. We adopt the following convention for the Fourier transform
\begin{equation}
\tilde{\mathcal{X}}(k^{\mu }) = \int d^{4}x\hspace{0.1cm}\exp \left( -ix_{\mu }k^{\mu}\right) \mathcal{X}(x^{\mu }), \ \ \ \
\mathcal{X}(x^{\mu }) = \int \frac{d^{4}k}{(2\pi )^{4}}\hspace{0.1cm}\exp \left(ix_{\mu }k^{\mu }\right) \tilde{\mathcal{X}}(k^{\mu }), \label{eq:def-fourier-trans}
\end{equation}
where $k^{\mu}=(\omega ,\mathbf{k})$, with $\omega$ being the frequency and $\mathbf{k}$ the wave vector. The next step is to express the linearized magneto-fluid-dynamical equations in Fourier space. For this purpose, it is convenient to introduce an orthonormal basis. Following Ref.~\cite{Brito:2020nou}, we first decompose the wave 4-vector as
\begin{equation}
k^{\mu }=\Omega u^\mu_0 +\kappa ^{\mu },
\end{equation}
where $\Omega =u^\mu_0 k_\mu$ and $\kappa^\mu = \Delta^{\mu \nu}_0 k_{\nu }$, with $\kappa ^{2}=-\kappa _{\mu }\kappa ^{\mu }$. We further decompose $\kappa^\mu$ in terms of its component parallel to the normalized magnetic 4-field and a component orthogonal to both $u^\mu_0$ and $b^\mu_0$,
\begin{equation}
\kappa ^{\mu }=\kappa_b b^\mu_0 +\kappa_\bot^\mu,
\end{equation}
with $\kappa_b = - b^\mu_0 \kappa_\mu$ and $\kappa_\bot^\mu = \Xi_{\nu,0}^\mu \kappa^\nu$. We finally define the 4-vector
\begin{equation}
\hat{q}^{\mu }=b_{0}^{\mu \nu}\hat{\kappa}_{\nu,\bot},
\end{equation}
where we introduced the normalized 4-vector, $\hat{\kappa}_{\bot }^\mu = \kappa_{\bot}^\mu /\kappa_{\bot } $, with $\kappa _{\bot }^{2}=\kappa ^{2}-\kappa _{b}^{2}$. We remark that $\hat{q}^\mu$ is orthogonal to $\hat{\kappa}_{\bot }^{\mu }$, $u^{\mu }$ and $b^{\mu }$. Employing the relation $b_{\left. {}\right. \alpha ,0}^{\lambda }b^{\alpha \beta}_0 = - \Xi ^{\lambda \beta }_0$, one can show that $\hat{q}^\mu$ is normalized, $\hat{q}_\mu \hat{q}^\mu =-1$. We also note that
\begin{equation}
b_{0}^{\nu\mu }\hat{q}_{\nu } = - \hat{\kappa}_{\bot }^{\mu }.
\end{equation}

We express the perturbations of the magneto-fluid-dynamical variables in terms of the orthonormal basis $\left\{ u_{0}^{\mu },b_{0}^{\mu },\hat{\kappa}_{\bot }^{\mu },\hat{q}^{\mu } \right\}$. Perturbations of the fluid 4-velocity in Fourier space become
\begin{equation}
\Delta \tilde{u}^\mu = \Delta \tilde{u}_b b^\mu_0 + \Delta \tilde{u}_k \hat{\kappa}_\bot^\mu + \Delta \tilde{u}_q \hat{q}^\mu,
\end{equation}
with $\Delta \tilde{u}_b = - b_\mu^0 \Delta \tilde{u}^\mu$, $\Delta \tilde{u}_k = -\hat{\kappa}_{\bot \mu} \Delta \tilde{u}^\mu$, and $\Delta \tilde{u}_q = - \hat{q}_\mu \Delta \tilde{u}^\mu$. Similarly, the normalized magnetic field can be decomposed as,
\begin{equation}
\Delta \tilde{b}^\mu = \Delta \tilde{b}_u u^\mu_0 + \Delta \tilde{b}_k \hat{\kappa}_\bot^\mu + \Delta \tilde{b}_q \hat{q}^\mu,
\end{equation}
where $\Delta \tilde{b}_u = u_\mu^0 \Delta \tilde{b}^\mu$, $\Delta \tilde{b}_k = -\hat{\kappa}_{\bot \mu} \Delta \tilde{b}^\mu$, and $\Delta \tilde{b}_q = - \hat{q}_\mu \Delta \tilde{b}^\mu$. The shear-stress tensor becomes
\begin{equation}
\Delta \tilde{\pi}^{\mu \nu }
= \Delta \tilde{\pi}_{bb} b_0^\mu b_0^\nu 
+ \Delta \tilde{\pi}_{kk} \hat{\kappa}_\bot^\mu \hat{\kappa}_\bot^\nu 
+ \Delta \tilde{\pi}_{qq} \hat{q}^\mu \hat{q}^\nu 
+ \Delta \tilde{\pi}_{bk} \left( \hat{\kappa}_\bot^\mu b_0^\nu + \hat{\kappa}_\bot^\nu b_0^\mu \right) 
+ \Delta \tilde{\pi}_{bq} \left( \hat{q}^\mu b_0^\nu +\hat{q}^\nu b_0^\mu
\right) 
+ \Delta \tilde{\pi}_{kq} \left( \hat{\kappa}_{\bot }^\mu \hat{q}^\nu + \hat{\kappa}_\bot^\mu \hat{q}^\nu \right),
\end{equation}
where each tensor component is defined as implied. The traceless property of the shear-stress tensor further implies that
\begin{equation}
\Delta \tilde{\pi}_{bb} + \Delta \tilde{\pi}_{kk} + \Delta \tilde{\pi}_{qq} = 0,
\end{equation}
leading to
\begin{align}
\label{eq:decomposition-shear}
\Delta \tilde{\pi}^{\mu \nu}
&= \Delta \tilde{\pi}_{bb} \left( b_0^\mu b_0^\nu + \frac{\Xi_0^{\mu\nu}}{2} \right) 
+ \left( \Delta \tilde{\pi}_{bb} + 2 \Delta \tilde{\pi}_{kk} \right) \left( \hat{\kappa}_\bot^\mu \hat{\kappa}_\bot^\nu + \frac{\Xi_0^{\mu\nu}}{2} \right) \notag \\
&+ \Delta \tilde{\pi}_{bk} \left( \hat{\kappa}_{\bot }^{\mu }b_{0}^{\nu }+\hat{\kappa}_{\bot }^{\nu }b_{0}^{\mu}\right) 
+ \Delta \tilde{\pi}_{bq}\left( \hat{q}^{\mu }b_{0}^{\nu }+\hat{q}^{\nu }b_{0}^{\mu}\right) 
+ \Delta \tilde{\pi}_{kq} \left( \hat{\kappa}_{\bot }^{\mu }\hat{q}^{\nu }+\hat{\kappa}_{\bot }^{\mu }\hat{q}^{\nu }\right).
\end{align}
An analogous expression can be written for the relative shear-stress tensor.

\subsection{Projected equations}
\label{subsec: Projected equations}

The next step is to project the linearized magneto-fluid-dynamical equations in Fourier space in terms of the basis introduced in the previous subsection. First, we express all the linearized equations, i.e., Eqs.~\eqref{eq:linear-cons-tmunu}--\eqref{eq:linearized-shear}, in Fourier space,
\begin{subequations}
\label{eq:linear-fourier}
\begin{align}
\left( \varepsilon _{0}+P_{0}+B_{0}^{2}\right) \Omega \Delta \tilde{u}^{\nu
}+B_{0}^{2}\kappa _{b}\Delta \tilde{b}^{\nu }+\left[ \left( \varepsilon
_{0}+P_{0}+B_{0}^{2}\right) \kappa _{\mu }\Delta \tilde{u}^{\mu }+\Omega
\left( \Delta \tilde{\varepsilon}+B_{0}\Delta \tilde{B}\right) \right]
u_{0}^{\nu }&- \notag \\
b_{0}^{\nu }\left[ \Omega B_{0}^{2}\Delta \tilde{u}_{b}+\kappa _{b}\left(
\Delta \tilde{P}-B_{0}\Delta \tilde{B}\right) -\kappa _{\bot
}B_{0}^{2}\Delta \tilde{b}_{k}\right] -\left( \Delta \tilde{P}+B_{0}\Delta 
\tilde{B}\right) \kappa _{\bot }^{\nu }+\kappa _{\mu }\Delta \tilde{\pi}%
^{\mu \nu }&=0,  \label{eq:fourier-cons-mom} \\
u^\mu_0 k_\nu \Delta \tilde{B}^\nu + B_0^\nu \kappa_\nu \Delta \tilde{u}^\mu - B_0^\mu \kappa_\nu \Delta \tilde{u}^\nu - \Omega \Delta \tilde{B}^\mu &= 0, \label{eq:fourier-maxwell-hodge} \\
\left( i\Omega + \Sigma_0 \right) \Delta \tilde{\pi}^{\mu \nu} + \omega_0 b_0^{\lambda \mu} \Delta \tilde{\delta \pi}_\lambda^\nu -\frac{4i}{15} \varepsilon_0 \left( \kappa ^{\mu } \Delta \tilde{u}^\nu + \kappa^\nu \Delta \tilde{u}^\mu - \frac{2}{3} \Delta^{\mu \nu} \kappa_\alpha \Delta \tilde{u}^\alpha \right) 
&= 0, \label{eq:fourier-total-shear} \\
\left( i\Omega + \Sigma_0' \right) \Delta \tilde{\delta \pi}^{\mu \nu} + \omega_0 b_0^{\lambda\mu} \Delta \tilde{\pi}_\lambda^\nu &= 0. \label{eq:fourier-relative-shear}
\end{align}
\end{subequations}

Equations \eqref{eq:fourier-cons-mom} and \eqref{eq:fourier-maxwell-hodge} are 4-vectors and can be decomposed with respect to our basis. In practice, this task is performed by contracting them with $u_\mu$ $b_\mu^0$, $\hat{\kappa}_{\mu, \bot}$ and $\hat{q}_\mu$. The Maxwell's equation for the Hodge dual then become,
\begin{subequations}
\begin{align}
\kappa_\bot \Delta \tilde{b}_k + \kappa_b \frac{\Delta \tilde{B}}{B_0} &= 0, \\
\Omega \frac{\Delta \tilde{B}}{B_0} - \kappa_\bot \Delta \tilde{u}_k &= 0, \\
\Omega \Delta \tilde{b}_k + \kappa_b \Delta \tilde{u}_k &= 0, \\
\Omega \Delta \tilde{b}_q + \kappa_b \Delta \tilde{u}_q &= 0.
\end{align}
\end{subequations}
The conservation law \eqref{eq:fourier-cons-mom}, further simplified using the Maxwell's equations above, take the following form,
\begin{subequations}
\label{eq:cons-laws-fourier-final}
\begin{align}
\Omega \Delta \tilde{\varepsilon} -\left( \varepsilon_0 + P_0 \right) \left( \kappa_b \Delta \tilde{u}_b + \kappa_\bot \Delta \tilde{u}_k \right) &= 0, \label{eq:cons-energ-final} \\
- \left( \varepsilon_0 + P_0 \right) \Omega \Delta \tilde{u}_b + \kappa_b \Delta \tilde{P} +\kappa_b \Delta \tilde{\pi}_{bb} + \kappa_\bot \Delta \tilde{\pi}_{bk} &= 0, \label{eq:cons-momentum-b} \\
\left[ -\left( \varepsilon_0 + P_0 + B_0^2 \right) \Omega^2 + B_0^2  \kappa^2  \right] \Delta \tilde{u}_k + \Omega \kappa_\bot \Delta \tilde{P} + \Omega \kappa_b \Delta \tilde{\pi}_{bk} + \Omega \kappa_\bot \Delta \tilde{\pi}_{kk} &= 0, \label{eq:cons-momentum-k} \\
\left[ -\left( \varepsilon_0 + P_0 + B_0^2 \right) \Omega^2 + B_0^2 \kappa_b^2 \right] \Delta \tilde{u}_q + \Omega \kappa_b \Delta \tilde{\pi}_{bq} + \Omega \kappa_\bot \Delta \tilde{\pi}_{kq} &= 0. \label{eq:cons-momentum-q}
\end{align}
\end{subequations}
Equations \eqref{eq:fourier-total-shear} and \eqref{eq:fourier-relative-shear} are symmetric, traceless second-rank tensors and will also be decomposed with respect to our basis. This is done by projecting them with all possible permutations of our basis elements:
\begin{itemize}
\item $\hat{\kappa}_{\bot \mu }\hat{\kappa}_{\bot \nu }$
\begin{subequations}
\label{eq:shear-kk}
\begin{align}
\left( i\Omega +\Sigma_0 \right) \Delta \tilde{\pi}_{kk} + \omega_0 \Delta \tilde{\delta \pi}_{kq} &=
\frac{8}{15} i \varepsilon_0 \left( \frac{2}{3}\kappa_\bot \Delta \tilde{u}_k - \frac{1}{3} \kappa_b \Delta \tilde{u}_b \right), \label{eq:total-shear-kk} \\
\left( i\Omega +\Sigma_0' \right) \Delta \tilde{\delta \pi}_{kk} + \omega_0 \Delta \tilde{\pi}_{kq} &= 0.
\end{align}
\end{subequations}

\item $b_\mu b_\nu$
\begin{subequations}
\label{eq:shear-bb}
\begin{align}
\left( i \Omega +\Sigma_0 \right) \Delta \tilde{\pi}_{bb} 
&= \frac{8}{15}i \varepsilon_0 \left( \frac{2}{3} \kappa_b \Delta \tilde{u}_b -\frac{1}{3} \kappa_\bot \Delta \tilde{u}_k \right), \label{eq:total-shear-bb} \\
\left( i\Omega + \Sigma_0' \right) \Delta \tilde{\delta \pi}_{bb} &= 0. \label{eq:relative-shear-bb}
\end{align}
\end{subequations}

\item $b_\mu \hat{\kappa}_{\nu, \bot}$
\begin{subequations}
\label{eq:shear-bk}
\begin{align}
\left( i\Omega +\Sigma_0 \right) \Delta \tilde{\pi}_{bk} + \frac{\omega_0}{2} \Delta \tilde{\delta \pi}_{bq} &=
\frac{4}{15} i \varepsilon_0 \left( \kappa_\bot \Delta \tilde{u}_b + \kappa _b \Delta \tilde{u}_k \right), \label{eq:total-shear-bk} \\
\left( i\Omega + \Sigma'_0 \right) \Delta \tilde{\delta \pi}_{bk} + \frac{\omega_0}{2} \Delta \tilde{\pi}_{bq} &= 0. \label{eq:relative-shear-bk}
\end{align}
\end{subequations}

\item $\hat{q}_\mu \hat{\kappa}_{\nu, \bot}$
\begin{subequations}
\label{eq:shear-qk}
\begin{align}
\left( i\Omega + \Sigma_0 \right) \Delta \tilde{\pi}_{kq} - \frac{\omega_0}{2} \left( \Delta \tilde{\delta \pi}_{bb} + 2 \Delta \tilde{\delta \pi}_{kk} \right) &=
\frac{4}{15} i \varepsilon_0 \kappa_\bot \Delta \tilde{u}_q, \label{eq:total-shear-qk} \\
\left( i\Omega +\Sigma'_{0} \right) \Delta \tilde{\delta \pi}_{kq} - \frac{\omega_0}{2} \left( \Delta \tilde{\pi}_{bb} + 2 \Delta \tilde{\pi}_{kk} \right) &= 0. \label{eq:relative-shear-qk}
\end{align}
\end{subequations}

\item $\hat{q}_\mu b_\nu $
\begin{subequations}
\label{eq:shear-bq}
\begin{align}
\left( i\Omega + \Sigma_0 \right) \Delta \tilde{\pi}_{bq} - \frac{\omega_0}{2} \Delta \tilde{\delta \pi}_{bk} &=
\frac{4}{15} i \varepsilon_0 \kappa_b \Delta \tilde{u}_q, \label{eq:total-shear-bq} \\
\left( i\Omega +\Sigma_0 \right) \Delta \tilde{\delta \pi}_{bq} - \frac{\omega_0}{2} \Delta \tilde{\pi}_{bk} &= 0. \label{eq:relative-shear-bq}
\end{align}
\end{subequations}
\end{itemize}

We remark that the equation for $\Delta \pi_{kq}$ couples with $\Delta \delta \pi_{bb}$ and $\Delta \delta \pi_{kk}$ via the term $\Delta \delta \pi_{bb} + 2 \Delta \delta \pi_{kk}$ (analogously, the equation for $\Delta \delta \pi_{kq}$ couples with $\Delta \pi_{bb} + 2 \Delta \pi_{kk}$). It is then convenient to define this quantity as an independent variable by itself, as was done in the decomposition of the shear-stress tensor in Eq.~\eqref{eq:decomposition-shear}, recasting the equation for $\Delta \delta \pi_{kk}$ in terms of this new variable. From Eqs.~\eqref{eq:shear-kk} and \eqref{eq:shear-bb}, we obtain
\begin{subequations}
\label{eq:shear-bb+2kk}
\begin{align}
\left( i \Omega + \Sigma_0 \right) \left( \Delta \tilde{\pi}_{bb} + 2 \Delta \tilde{\pi}_{kk} \right) + 2 \omega_0 \Delta \tilde{\delta \pi}_{kq} &=
\frac{8}{15} i \varepsilon_0 \kappa_\bot \Delta \tilde{u}_k, \label{eq:total-shear-bb+2kk} \\
\left( i \Omega + \Sigma_0' \right) \left( \Delta \tilde{\delta \pi}_{bb} + 2 \Delta \tilde{\delta \pi}_{kk} \right) + 2 \omega_0 \Delta \tilde{\pi}_{kq} &= 0. \label{eq:relative-shear-bb+2kk}
\end{align}
\end{subequations}

Finally, it is convenient to eliminate all terms containing the relative shear-stress tensor, $\Delta \delta \pi$, from Eqs.~\eqref{eq:total-shear-bk}--\eqref{eq:total-shear-bb+2kk} using Eqs.~\eqref{eq:relative-shear-bk}--\eqref{eq:relative-shear-bb+2kk}. We then obtain,
\begin{subequations}
\begin{align}
\left( i\Omega +\Sigma_0 + \frac{\omega_0^2}{i \Omega + \Sigma'} \right) \left( \Delta \tilde{\pi}_{bb} + 2 \Delta \tilde{\pi}_{kk} \right)
&= \frac{8}{15} i \varepsilon_0 \kappa_\bot \Delta \tilde{u}_k, \label{eq:total-bb+2kk-final} \\
\left[ i \Omega + \Sigma_0 + \frac{\omega_0^2}{4 (i \Omega + \Sigma')} \right] \Delta \tilde{\pi}_{bk}
&= \frac{4}{15} i \varepsilon_0 \left( \kappa_\bot \Delta \tilde{u}_b + \kappa_b \Delta \tilde{u}_k \right), \label{eq:total-bk-final}  \\
\left( i\Omega +\Sigma_0 + \frac{\omega_0^2}{i \Omega + \Sigma'} \right) \Delta \tilde{\pi}_{kq} 
&= \frac{4}{15} i \varepsilon_0 \kappa_\bot \Delta \tilde{u}_q, \label{eq:total-kq-final}  \\
\left[ i\Omega +\Sigma_0 + \frac{\omega_0^2}{4 (i \Omega + \Sigma')} \right] \Delta \tilde{\pi}_{bq}  
&= \frac{4}{15} i \varepsilon_0 \kappa_b \Delta \tilde{u}_q. \label{eq:total-bq-final} 
\end{align}
\end{subequations}
We remark that the system of equations above partially decouples, with
Eqs.~\eqref{eq:cons-energ-final}, \eqref{eq:cons-momentum-b}, \eqref{eq:cons-momentum-k}, \eqref{eq:total-shear-bb}, \eqref{eq:total-bb+2kk-final} and \eqref{eq:total-bk-final} being solved independently from Eqs.~\eqref{eq:cons-momentum-q}, \eqref{eq:total-kq-final}, and \eqref{eq:total-bq-final}. Finally, we note that equation \eqref{eq:relative-shear-bb} can be solved directly, leading to the trivial nonhydrodynamic mode $\Omega = i\Sigma'_0$, and will not be discussed further.

\subsection{Rescaled linear magneto-fluid-dynamical equations in Fourier space}
\label{subsec:resc-linear-mhd-four}

For the sake of convenience, we define the dimensionless variables,
\begin{equation}
\Delta \hat{\varepsilon} = \frac{\Delta \tilde{\varepsilon}}{\varepsilon_0 + P_0}, \quad
\Delta \hat{\pi} = \frac{\Delta \tilde{\pi}}{\varepsilon_0 + P_0}.
\end{equation}
Then, using that $P_0 = \varepsilon_0 /3 $, we rewrite Eqs.~\eqref{eq:cons-energ-final}--\eqref{eq:cons-momentum-k}, \eqref{eq:total-shear-bb}, \eqref{eq:total-bb+2kk-final} and \eqref{eq:total-bk-final} as 
\begin{subequations}
\label{eq:bk-components-final}
\begin{align}
\Omega \Delta \hat{\varepsilon} - \kappa_b \Delta \tilde{u}_b - \kappa_\bot \Delta \tilde{u}_k 
&= 0, \label{eq:cons-energ-final-rescaled} \\
- \Omega \Delta \tilde{u}_b + \frac{\kappa_b}{3} \Delta \hat{\varepsilon} + \kappa_b \Delta \hat{\pi}_{bb} + \kappa_\bot \Delta \hat{\pi}_{bk} 
&= 0, \label{eq:cons-momentum-b-rescaled} \\
\left( i \Omega +\Sigma_0 \right) \Delta \hat{\pi}_{bb} - \frac{2i}{15} \left( 2 \kappa_b \Delta \tilde{u}_b - \kappa_\bot \Delta \tilde{u}_k \right) 
&= 0, \label{eq:total-shear-bb-rescaled} \\
\left[ - ( 1 + \mathcal{B} ) \Omega^2 + \mathcal{B} \left( \kappa^2_b + \kappa^2_\bot \right) \right] \Delta \tilde{u}_k + \Omega \frac{\kappa_\bot}{3} \Delta \hat{\varepsilon} + \Omega \kappa_b \Delta \hat{\pi}_{bk} + \Omega \kappa_\bot \Delta \hat{\pi}_{kk} 
&= 0, \label{eq:cons-momentum-k-rescaled} \\
\left( i \Omega + \Sigma_0 + \frac{\omega_0^2}{i \Omega + \Sigma'} \right) \left( \Delta \hat{\pi}_{bb} + 2 \Delta \hat{\pi}_{kk} \right) - \frac{2 i}{5} \kappa_\bot \Delta \tilde{u}_k 
&= 0, \label{eq:total-bb+2kk-final-rescaled} \\
\left[ i\Omega +\Sigma_0 + \frac{\omega_0^2}{4 (i \Omega + \Sigma')} \right] \Delta \hat{\pi}_{bk}
- \frac{i}{5} \left( \kappa_\bot \Delta \tilde{u}_b + \kappa_b \Delta \tilde{u}_k \right) &= 0, \label{eq:total-bk-final-rescaled}
\end{align}
\end{subequations}
where we have defined 
\begin{equation}
\mathcal{B} = \frac{B_0^2}{\varepsilon_0 + P_0}.
\end{equation}

Similarly, rescaling Eqs.~\eqref{eq:cons-momentum-q}, \eqref{eq:total-kq-final} and \eqref{eq:total-bq-final}, we obtain
\begin{subequations}
\label{eq:q-components-final}
\begin{align}
\left[ - ( 1 + \mathcal{B} ) \Omega^2 + \mathcal{B} \kappa^2_b \right] \Delta \tilde{u}_q + \Omega \kappa_b \Delta \hat{\pi}_{bq} + \Omega \kappa_\bot \Delta \hat{\pi}_{kq} 
&= 0, \label{eq:cons-mom-q-final} \\
\left( i\Omega + \Sigma_0 + \frac{\omega_0^2}{i \Omega + \Sigma'} \right) \Delta \hat{\pi}_{kq} - \frac{i}{5} \kappa_\bot \Delta \tilde{u}_q &= 0,  \label{eq:total-kq-rescaled-final}\\
\left[ i\Omega + \Sigma_0 + \frac{\omega_0^2}{4 (i \Omega + \Sigma')} \right] \Delta \hat{\pi}_{bq} - \frac{i}{5} \kappa_b \Delta \tilde{u}_q 
&= 0. \label{eq:total-bq-rescaled-final}
\end{align}
\end{subequations}

\section{Causality and stability analysis}
\label{sec:stab-analysis}

We can now determine the dispersion relations resulting from Eqs.~\eqref{eq:bk-components-final} and \eqref{eq:q-components-final} and verify whether the magneto-fluid-dynamical theory developed in Ref.~\cite{Kushwah:2024zgd} is linearly causal and stable. Linear stability is ensured as long as the modes have a positive imaginary part, whereas linear causality is guaranteed if the asymptotic group velocity is smaller than the speed of light,
\begin{equation}
\label{eq:causality-condition}
\lim_{k \rightarrow \infty} \left\vert \frac{\partial \, \mathrm{Re} (\omega)}{\partial k} \right\vert \leq 1.
\end{equation}
In the following analyses we shall consider two different scenarios: (i) in the first we assume that the perturbations are longitudinal with respect to the magnetic field,
$\kappa^{\mu} \ \| \ b_0^\mu$, which implies that $\kappa_\bot = 0$, while, (ii) in the second case we assume that the perturbations are transverse to the magnetic field, $\kappa^{\mu} \perp  b_0^\mu$, which implies that $\kappa_b = 0$. 

Furthermore, it was shown in Ref.~\cite{Gavassino:2021owo} that if linear causality and stability are satisfied for perturbations of a fluid in a static background, then they are also satisfied for perturbations of a fluid in a moving background. Therefore, in what follows, we restrict ourselves to perturbations on a static fluid,
\begin{equation}
u^\mu_0 = (1,0,0,0),
\end{equation}
which yields considerably simpler dispersion relations. Nevertheless, for the sake of completeness, we first  obtain the dispersion relations in their most general form, and only then impose the aforementioned restrictions.

\subsection{Longitudinal perturbations}
\label{subsec:long-pert-full-theory}

We first consider longitudinal perturbations with respect to the background magnetic field, i.e., we assume that $\kappa_\bot = 0$. Then, Eqs.~\eqref{eq:cons-energ-final-rescaled}--\eqref{eq:total-shear-bb-rescaled} decouple from Eqs.~\eqref{eq:cons-momentum-k-rescaled}--\eqref{eq:total-bk-final-rescaled} and reduce to 
\begin{subequations}
\label{eq:b-components-k-ortho-b}
\begin{align}
\Omega \Delta \hat{\varepsilon} - \kappa_b \Delta \tilde{u}_b &= 0, \\
- \Omega \Delta \tilde{u}_b + \frac{\kappa_b}{3} \Delta \hat{\varepsilon} + \kappa_b \Delta \hat{\pi}_{bb} &= 0, \\
\left( i \Omega +\Sigma_0 \right) \Delta \hat{\pi}_{bb} - \frac{4i}{15} \kappa_b \Delta \tilde{u}_b &= 0. 
\end{align}
\end{subequations}
Furthermore, Eqs.~\eqref{eq:cons-momentum-k-rescaled}--\eqref{eq:total-bk-final-rescaled} become,
\begin{subequations}
\label{eq:k-components-k-ortho-b}
\begin{align}
\left[ - ( 1 + \mathcal{B} ) \Omega^2 + \mathcal{B}\kappa^2_b \right] \Delta \tilde{u}_k +  \Omega \kappa_b \Delta \hat{\pi}_{bk} &= 0, \\
\left( i \Omega + \Sigma_0 + \frac{\omega_0^2}{i \Omega + \Sigma'} \right) \left( \Delta \hat{\pi}_{bb} + 2 \Delta \hat{\pi}_{kk} \right) &= 0, \\
\left[ i\Omega +\Sigma_0 + \frac{\omega_0^2}{4 (i \Omega + \Sigma')} \right] \Delta \hat{\pi}_{bk}
- \frac{i}{5} \kappa_b \Delta \tilde{u}_k &= 0. 
\end{align}
\end{subequations}
Finally, Eqs.~\eqref{eq:q-components-final} simplify to
\begin{subequations}
\label{eq:q-components-k-ortho-b}
\begin{align}
\left[ - ( 1 + \mathcal{B} ) \Omega^2 + \mathcal{B} \kappa^2_b \right] \Delta \tilde{u}_q + \Omega \kappa_b \Delta \hat{\pi}_{bq}
&= 0, \\
\left( i\Omega + \Sigma_0 + \frac{\omega_0^2}{i \Omega + \Sigma'} \right) \Delta \hat{\pi}_{kq} &= 0, \\
\left[ i\Omega + \Sigma_0 + \frac{\omega_0^2}{4 (i \Omega + \Sigma')} \right] \Delta \hat{\pi}_{bq} - \frac{i}{5} \kappa_b \Delta \tilde{u}_q 
&= 0. 
\end{align}
\end{subequations}
In particular, note that Eqs.~\eqref{eq:k-components-k-ortho-b} and \eqref{eq:q-components-k-ortho-b} lead to the same dispersion relations and, thus, to a set of identical solutions, i.e., they yield degenerate modes. As previously mentioned, we consider perturbations on a plasma at rest, which implies that $\Omega = \omega$ and $\kappa_b = k$. Then, from Eqs.~\eqref{eq:b-components-k-ortho-b} and \eqref{eq:k-components-k-ortho-b} [or, equivalently, Eqs.~\eqref{eq:b-components-k-ortho-b} and \eqref{eq:q-components-k-ortho-b}], we obtain the following dispersion relations
\begin{subequations}
\label{eq:long-modes-general}
\begin{align}
\omega^2 - i (\Sigma + \Sigma') \omega - \Sigma \Sigma' - \omega_0^2 &= 0, \label{eq:long-disp-rel-exact-solvable} \\
\omega^3 - i \Sigma \omega^2 - \frac{3}{5} k^2 \omega + i \frac{\Sigma}{3} k^2 &= 0, \label{eq:long-B-independent-modes} \\
\omega^4 - i (\Sigma + \Sigma') \omega^3 - \left[ \frac{1 + 5 \mathcal{B}}{5 (1 + \mathcal{B})} k^2 + \Sigma \Sigma' + \frac{\omega_0^2}{4} \right] \omega^2 + i \left( \frac{1 + 5 \mathcal{B}}{5 (1 + \mathcal{B})} \Sigma' + \frac{\mathcal{B}}{1 + \mathcal{B}} \Sigma \right) k^2 \omega + \frac{\mathcal{B} k^2}{1 + \mathcal{B}} \left( \frac{\omega_0^2}{4} + \Sigma \Sigma' \right) &=0. \label{eq:long-B-dependent-modes}
\end{align}
\end{subequations}
The dispersion relation \eqref{eq:long-B-independent-modes} does not display any dependence on the magnetic field and is identical to the dispersion relation obtained for the sound channel \cite{Brito:2020nou} in the usual Israel-Stewart theory --  in this case, one identifies the variable $\Sigma$ as the inverse shear relaxation time. The remaining dispersion relations depend significantly on the magnetic field, through the quantities $\mathcal{B}$ and $\omega_0$.

We remark that Eq.~\eqref{eq:long-disp-rel-exact-solvable} is the only dispersion above that leads to simple analytic solutions,
\begin{equation}
\label{eq:exact-longitudinal-modes}
\omega_\pm = i \frac{\Sigma + \Sigma' \pm \sqrt{(\Sigma - \Sigma')^2 - 4 \omega_0^2}}{2}.
\end{equation}
Since $\Sigma$ and $\Sigma'$ are positive-definite transport coefficients, it can be readily seen that these modes are always stable. Furthermore, for sufficiently large values of magnetic field (which is encoded in $\omega_0$), both solutions have non-zero real parts, and thus become oscillating as well as damping, as was also observed in Ref.~\cite{Kushwah:2024zgd}. In this case, the imaginary parts of the modes no longer display a dependence on the value of $\omega_0$. 

The remaining dispersion relations do not have simple analytical solutions and we thus restrict our analyses to the asymptotic behavior of the modes, i.e., in the small ($k \to 0$) and large ($k \to \infty$) wavenumber limits. For small $k$, the longitudinal modes read
\begin{subequations}
\begin{align}
\omega_{\mathrm{nh}} (k) &= i \Sigma - \frac{4 i}{15 \Sigma} k^2 + \mathcal{O} (k^3), \label{eq:usual0} \\
\omega_{\mathrm{nh}} (k) &= \frac{i}{2} \left[ \Sigma + \Sigma' \pm \sqrt{( \Sigma - \Sigma')^2 - \omega_0^2} \right] + \mathcal{O} (k^2), \label{eq:bla} \\
\omega_\mathrm{h} (k) &= \pm \frac{1}{\sqrt{3}} k + \frac{2i}{15 \Sigma} k^2 + \mathcal{O} (k^3), \label{eq:usual1} \\
\omega_\mathrm{h} (k) &= \pm v_{\mathrm{A}} k + \frac{2i}{5} \frac{\Sigma'}{(1+ \mathcal{B}) \left( 4 \Sigma \Sigma' + \omega_0^2 \right)} k^2 + \mathcal{O} (k^3), \label{eq:alfvenmode}
\end{align}
\end{subequations}
where we have defined the Alfvén velocity as $v_{\mathrm{A}} = \sqrt{\frac{\mathcal{B}}{1+ \mathcal{B}}}$ \cite{jackson2012classical}. The subscripts `nh' and `h' stand for non-hydrodynamic and hydrodynamic modes, with the latter denoting solutions that vanish at zero wavenumber and the former denoting modes that are gapped. The hydrodynamic mode \eqref{eq:usual1} describes the propagation of usual sound waves, with a velocity of sound $c_s^2=1/3$, and a diffusion-like damping term $\sim i k^2/\Sigma$. The nonhydrodynamic mode \eqref{eq:usual0} describes the damping ($\sim i \Sigma$) of the longitudinal component of the shear-stress tensor. These modes arise from dispersion relation \eqref{eq:long-B-independent-modes} and, as already mentioned, do not depend on the magnetic field. The hydrodynamic mode \eqref{eq:alfvenmode} describes the propagation of waves with the Alfvén velocity and their diffusion-like damping term, which displays a strong dependence on the value of the magnetic field. As a matter of fact, when the magnetic field becomes large, $B_0 \gg T_0^2$, the Alfvén velocity tends to $v_A \to 1$, while the damping term behaves as $\sim i \Sigma' T_0^6 k^2/ B_0^4$, becoming parametrically small. Finally, the nonhydrodynamic mode \eqref{eq:bla} describes the damping of the transverse perturbations of the shear-stress tensor. In this case, when the magnetic field is sufficiently large, one can see that these otherwise purely imaginary non-hydrodynamic modes can also display a non-zero real part -- that is, they become oscillating as well as damping modes \cite{Kushwah:2024zgd}. Most importantly, all modes have positive-definite imaginary parts and, therefore, these modes of the theory are identically stable, at least in the small wavenumber limit.

In the large wavenumber limit, the modes become
\begin{subequations}
\begin{align}
\omega (k) &= \pm \sqrt{\frac{3}{5}} k + \frac{2\Sigma i}{9} + \mathcal{O} \left( \frac{1}{k} \right), \\
\omega (k) &= \frac{5 \Sigma i}{9} + \frac{500 \Sigma^3 i}{2187} \frac{1}{k^2} + \mathcal{O} \left( \frac{1}{k^3} \right), \\
\omega (k) &= \pm \sqrt{\frac{1 + 5 \mathcal{B}}{5 + 5 \mathcal
B}} k + \frac{i \Sigma}{2 (1 + 5 \mathcal{B})} + \mathcal{O} \left( \frac{1}{k} \right), \\
\omega (k) &= \frac{i}{2 (1 + 5 \mathcal{B})} \left[ \Sigma' + 5 \mathcal{B} ( \Sigma' + \Sigma ) \pm \sqrt{[\Sigma' + 5 \mathcal{B} ( \Sigma' - \Sigma )]^2 - 5 \mathcal{B} (1 + 5 \mathcal{B}) \omega_0^2} \right] + \mathcal{O} \left( \frac{1}{k^2} \right).
\end{align}
\end{subequations}
The condition for linear causality \eqref{eq:causality-condition} is satisfied for all the above modes, for any value of magnetic field and cross section (contained in the positive-definite parameters $\Sigma$ and $\Sigma'$). We also note that all modes remain stable in this asymptotic limit, regardless of the values of magnetic field and cross sections, as expected of a self-consistent theory. In this paper, we do not obtain causality conditions for transport coefficients since these are fixed and were determined from the microscopic theory \cite{Kushwah:2024zgd}. 

\begin{figure}[ht]%[!ht]
\centering
\includegraphics[width=0.9\linewidth]{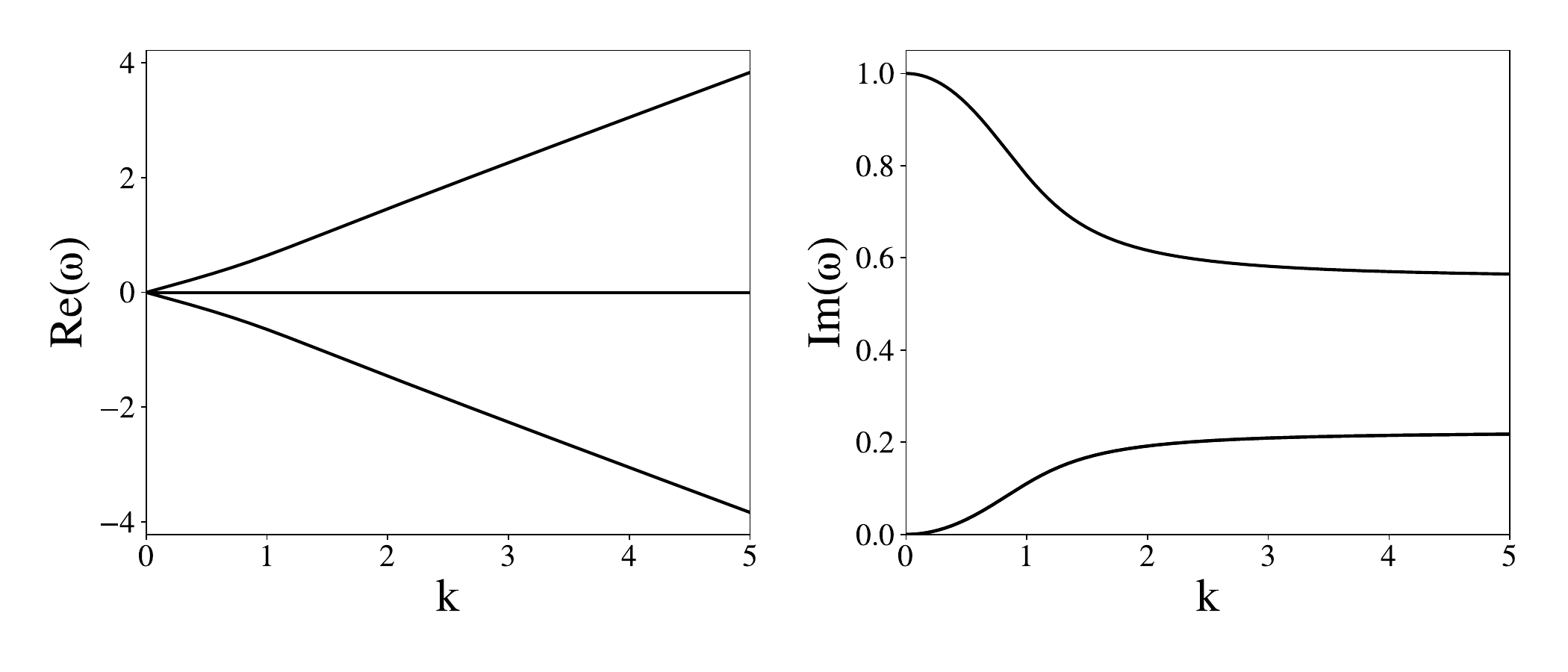} 
\caption{Solutions of Eq.~\eqref{eq:long-B-independent-modes} for $\Sigma = 1$.}
\label{fig:long-B-indep}
\end{figure}

We now plot the solutions for the dispersion relations for intermediate values of wavenumber. The solutions of Eq.~\eqref{eq:long-B-independent-modes} are portrayed as function of the wavenumber in Fig.~\ref{fig:long-B-indep} for $\Sigma = 1$. These are the only modes that do not possess any dependence on the magnetic field and, also, on the quantity $\Sigma'$ (the coupling between the total shear-stress tensor and the relative one effectively disappears). We remark that different values of $\Sigma$ do not yield appreciable qualitative changes in the modes -- in particular, larger (smaller) values of $\Sigma$ simply lead to modes that saturate slower (faster). In any case, linear causality and stability are always satisfied.  

\begin{figure}[ht]%[!ht]
\centering
\includegraphics[width=0.9\linewidth]{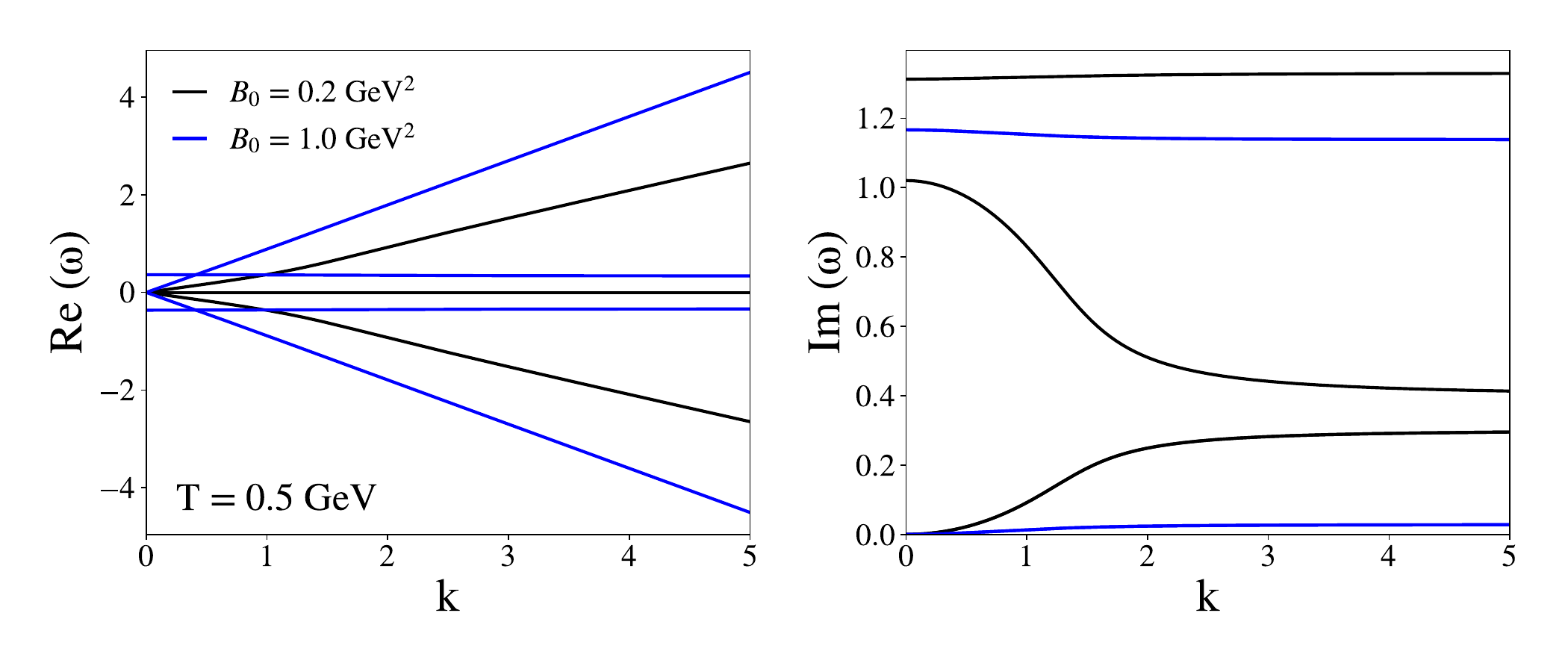} \\
\includegraphics[width=0.9\linewidth]{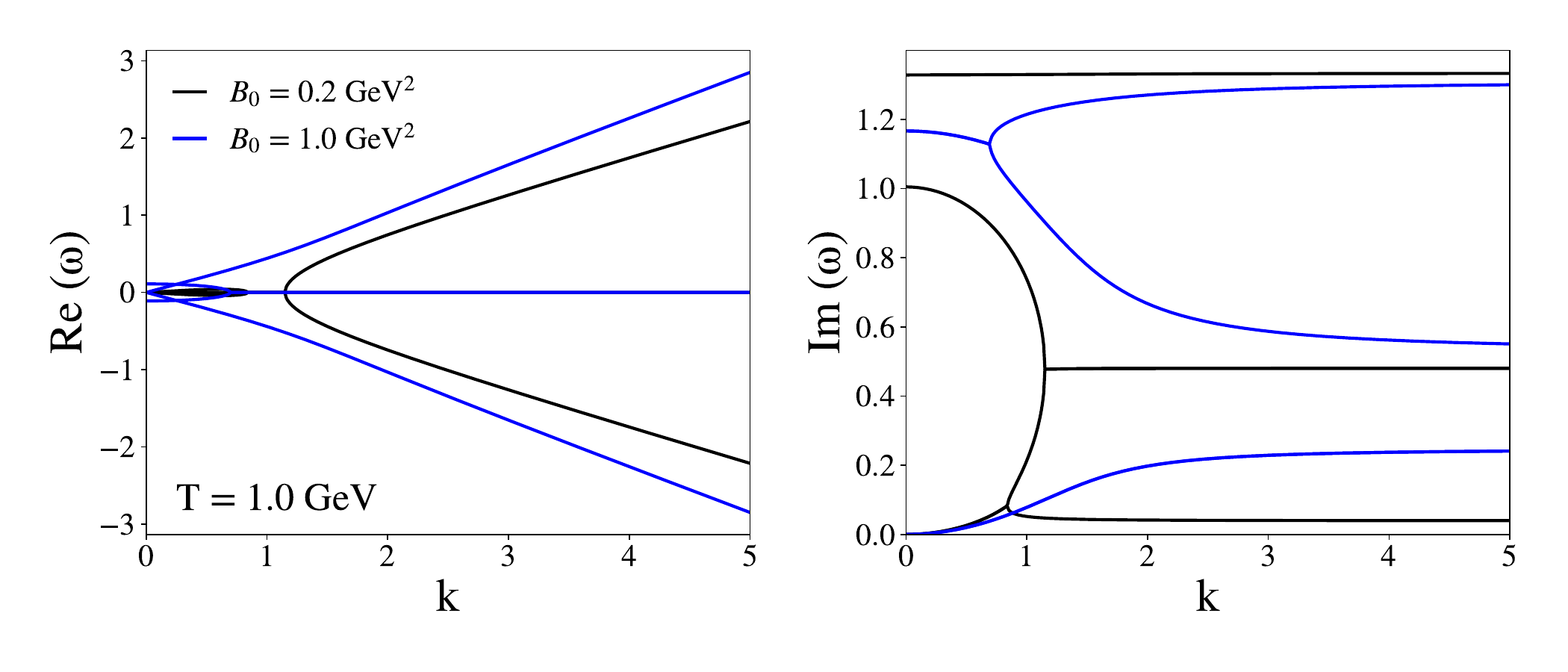}
\caption{Solutions of Eq.~\eqref{eq:long-B-dependent-modes} for $\Sigma = 1$, $\Sigma' = 4 \Sigma/3$ considering $B_0 = 0.2$ GeV$^2$ (black lines) and $B_0 = 1$ GeV$^2$ (blue lines) and $T = 0.5$ GeV (upper panels) and $T = 1$ GeV (lower panels).}
\label{fig:long-B-dep}
\end{figure}

In Fig.~\ref{fig:long-B-dep}, we display the solutions of Eq.~\eqref{eq:long-B-dependent-modes} for different values of $B_0$ and temperature, considering $\Sigma = 1$ and $\Sigma' = 4 \Sigma/3 $. We see that the modes display a significant dependence on the value of the magnetic field, which can trigger the emergence of oscillatory nonhydrodynamic modes. In particular, in the lower panel of Fig.~\ref{fig:long-B-dep}, we see that large values of $B_0$ lead to larger oscillation frequencies, as already hinted by our small wavenumber expansion. The imaginary part of the modes display a significant dependence on the value of magnetic field, with the hydrodynamic mode becoming almost non-dissipative for large values of $B_0/T_0^2$. 

\subsection{Transverse perturbations}
\label{subsec: Transverse modes}

We now consider the case of perturbations that are transverse to the background magnetic field, i.e., we assume $\kappa_b = 0$. As before, we consider perturbations on a plasma at rest, which here implies that $\Omega = \omega$ and $\kappa_\bot = k$.
Then, Eqs.~\eqref{eq:cons-energ-final-rescaled}, \eqref{eq:total-shear-bb-rescaled}, \eqref{eq:cons-momentum-k-rescaled}, \eqref{eq:total-bb+2kk-final-rescaled} decouple from Eqs.~\eqref{eq:cons-momentum-b-rescaled}, \eqref{eq:total-bk-final-rescaled}, resulting in two independent sets of equations of motion, namely
\begin{subequations}
\label{eq:coupled-eqs-case2-1}
\begin{align}
\omega \Delta \hat{\varepsilon} - k \Delta \tilde{u}_{k} &= 0, \\
\left( i \omega + \Sigma_0 \right) \Delta \hat{\pi}_{bb} + \frac{2i}{15} k \Delta \tilde{u}_k &= 0, \\
\left[ - ( 1 + \mathcal{B} ) \omega^2 + \mathcal{B} k^2_\bot \right] \Delta \tilde{u}_k + \omega \frac{k}{3} \Delta \hat{\varepsilon} + \frac{1}{2} \omega k \left( \Delta \hat{\pi}_{bb} + 2 \Delta \hat{\pi}_{kk} \right) - \frac{1}{2} \omega k \Delta \hat{\pi}_{bb}
&= 0, \\
\left( i \omega + \Sigma_0 + \frac{\omega_0^2}{i \omega + \Sigma'} \right) \left( \Delta \hat{\pi}_{bb} + 2 \Delta \hat{\pi}_{kk} \right) - \frac{2 i}{5} k \Delta \tilde{u}_k 
&= 0, 
\end{align}
\end{subequations}
and
\begin{subequations}
\label{eq:coupled-eqs-case2-2}
\begin{align}
\omega \Delta \tilde{u}_b - k \Delta \hat{\pi}_{bk} &= 0, \\
\left[ i\omega +\Sigma_0 + \frac{\omega_0^2}{4 (i \omega + \Sigma')} \right] \Delta \hat{\pi}_{bk}
- \frac{i}{5} k \Delta \tilde{u}_b &= 0.
\end{align}
\end{subequations}
Furthermore, Eqs.~\eqref{eq:q-components-final} simplify to
\begin{subequations}
\label{eq:q-components-trans}
\begin{align}
\left(  1 + \mathcal{B} \right) \omega^2 \Delta \tilde{u}_q - \omega k \Delta \hat{\pi}_{kq} &= 0,  \\
\left( i\omega + \Sigma_0 + \frac{\omega_0^2}{i \omega + \Sigma'} \right) \Delta \hat{\pi}_{kq} - \frac{i}{5} k \Delta \tilde{u}_q &= 0,  \\
\left[ i\omega + \Sigma_0 + \frac{\omega_0^2}{4 (i \omega + \Sigma')} \right] \Delta \hat{\pi}_{bq} 
&= 0. 
\end{align}
\end{subequations}
The above equations lead to the following set of dispersion relations
\begin{subequations}
\label{eq:disp-real-trans}
\begin{align}
(i \omega + \Sigma') (i\omega + \Sigma) + \frac{\omega_0^2}{4} &= 0, \label{eq:trans-disp-rel-q-components} \\
\omega^2 \left\{ \left[ k^2 \left[ 5 \Sigma  (3 \mathcal{B}+1)+3 i \omega  (5 \mathcal{B}+2) \right] - 15 \omega ^2 (\mathcal{B}+1) (\Sigma +i \omega )\right] \left[ \omega_0^2+(\Sigma +i \omega ) \left(\Sigma '+i \omega \right)\right] \right. \notag \\
- \left. 3 k^2 \omega  (\omega -i \Sigma ) \left(\Sigma '+i \omega \right) \right\} &= 0, \label{eq:trans-disp-rel-1} \\
\omega^3 - i \left( \Sigma + \Sigma' \right) \omega^2 - \left( \Sigma \Sigma' + \frac{k}{5} + \frac{\omega_0^2}{4} \right) \omega + \frac{\Sigma'}{5} i k^2 &= 0, \label{eq:trans-disp-rel-2} \\
\omega^4 - i (\Sigma + \Sigma') \omega^3 - \left[ \frac{1}{5 (1 + \mathcal{B})} k^2 + \Sigma \Sigma' + \omega_0^2 \right] \omega^2 + i \frac{1}{5 (1 + \mathcal{B})} \Sigma k^2 \omega &=0. \label{eq:pert-ub-plane-disprel-bk}
\end{align}
\end{subequations}
It can immediately be seen the existence of three trivial modes, $\omega = 0$. Unlike what was observed for the dispersion relations arising from longitudinal perturbations, all dispersion relations associated with transverse perturbations are affected by the magnetic field. We note that Eq.~\eqref{eq:trans-disp-rel-q-components} is the only dispersion relation that admits simple analytical solutions, given by
\begin{equation}
\omega_\pm = i \frac{\Sigma + \Sigma' \pm \sqrt{(\Sigma - \Sigma')^2 - \omega_0^2}}{2}.
\end{equation}
This expression closely resembles Eq.~\eqref{eq:exact-longitudinal-modes}, with the replacement $\omega_0 \to 2\omega_0$. Consequently, following the same reasoning as before, we conclude that these modes are stable. Furthermore, in the limit of sufficiently large magnetic fields, these otherwise purely imaginary modes acquire a non-zero real part, introducing an oscillatory dynamics along with damping. 

The remaining dispersion relations listed in Eqs.~\eqref{eq:disp-real-trans} do not admit simple analytical solutions. Therefore, as done in the previous subsection, these modes will be analyzed solely in the asymptotic limits $k \to 0$ and $k \to \infty$. In the small wavenumber regime, the modes take the following form
\begin{subequations}
\begin{align}
\omega_{\mathrm{nh}} &= i \Sigma + \mathcal{O} (k^2), \label{eq:trans-mode-nh1} \\
\omega_{\mathrm{nh}} &= \frac{i}{2} \left( \Sigma + \Sigma' \pm \sqrt{(\Sigma - \Sigma')^2 - 4 \omega_0^2} \right) + \mathcal{O} (k^2), \label{eq:trans-mode-nh2} \\
\omega_{\mathrm{nh}} &= \frac{i}{2} \left( \Sigma + \Sigma' \pm \sqrt{(\Sigma - \Sigma')^2 - \omega_0^2} \right) + \mathcal{O} (k^2), \label{eq:trans-mode-nh3} \\
\omega_{\mathrm{h}} &= \pm \sqrt{\frac{1 + 3 \mathcal{B}}{3 (1 + \mathcal{B})}} k + \frac{i}{30 (1 + \mathcal{B})} \frac{4 \Sigma \Sigma' + \omega_0^2}{\Sigma (\Sigma \Sigma' + \omega_0^2)} k^2 + \mathcal{O} (k^2), \label{eq:trans-mode-sound} \\
\omega_{\mathrm{h}} &= \frac{4 i}{5} \frac{\Sigma'}{4 \Sigma \Sigma' + \omega_0^2} k^2 + \mathcal{O} (k^3), \label{eq:trans-mode-hydro-1} \\
\omega_{\mathrm{h}} &= \frac{i}{10 (1 + \mathcal{B})} \frac{\Sigma}{\Sigma \Sigma' + \omega_0^2} k^2 +\mathcal{O} \left( k^3 \right). \label{eq:trans-mode-hydro-2}
\end{align}
\end{subequations}
The nonhydrodynamic modes \eqref{eq:trans-mode-nh1}--\eqref{eq:trans-mode-nh3} closely resemble their longitudinal counterparts \eqref{eq:usual0}, \eqref{eq:bla} and \eqref{eq:exact-longitudinal-modes}, respectively. These modes have been thoroughly discussed in the previous subsection and thus shall not be revisited here. As before, the hydrodynamic modes describe the propagation of sound waves or the diffusion of hydrodynamic perturbations. The modes \eqref{eq:trans-mode-sound} describe sound propagation, with a velocity of sound that depends on the magnetic field $\sqrt{\frac{1 + 3 \mathcal{B}}{3 (1 + \mathcal{B})}}$, and a dissipative term $\sim ik^2$ that also depends on the magnetic field (and cross sections). In particular, when the magnetic field is very large, this magnetic-field-dependent sound velocity goes to 1 and the diffusion-like term behaves as $\sim i T_0^4 k^2/ (\Sigma B_0^2)$, hence displaying a slower suppression as compared to the Alfv\'en mode \eqref{eq:alfvenmode}. In addition, in this limiting case this mode depends on the cross section via $\sim 1/\Sigma$ while the Alfv\'en mode depends on the cross section via $\sim \Sigma'$, implying that for weakly collisional plasmas ($\Sigma \to 0$ and $\Sigma' \to 0$) the former mode can become considerably more dissipative than the latter. The hydrodynamic modes \eqref{eq:trans-mode-hydro-1} and \eqref{eq:trans-mode-hydro-2} are purely diffusive and, at large values of magnetic field behave as $\sim i \Sigma' T_0^4 k^2/B_0^2$ and $\sim i \Sigma T_0^6 k^2/B_0^4$, respectively. Overall, in the small wavenumber limit all these modes are stable.

\begin{figure}[ht]%[!ht]
\centering
\includegraphics[width=0.9\linewidth]{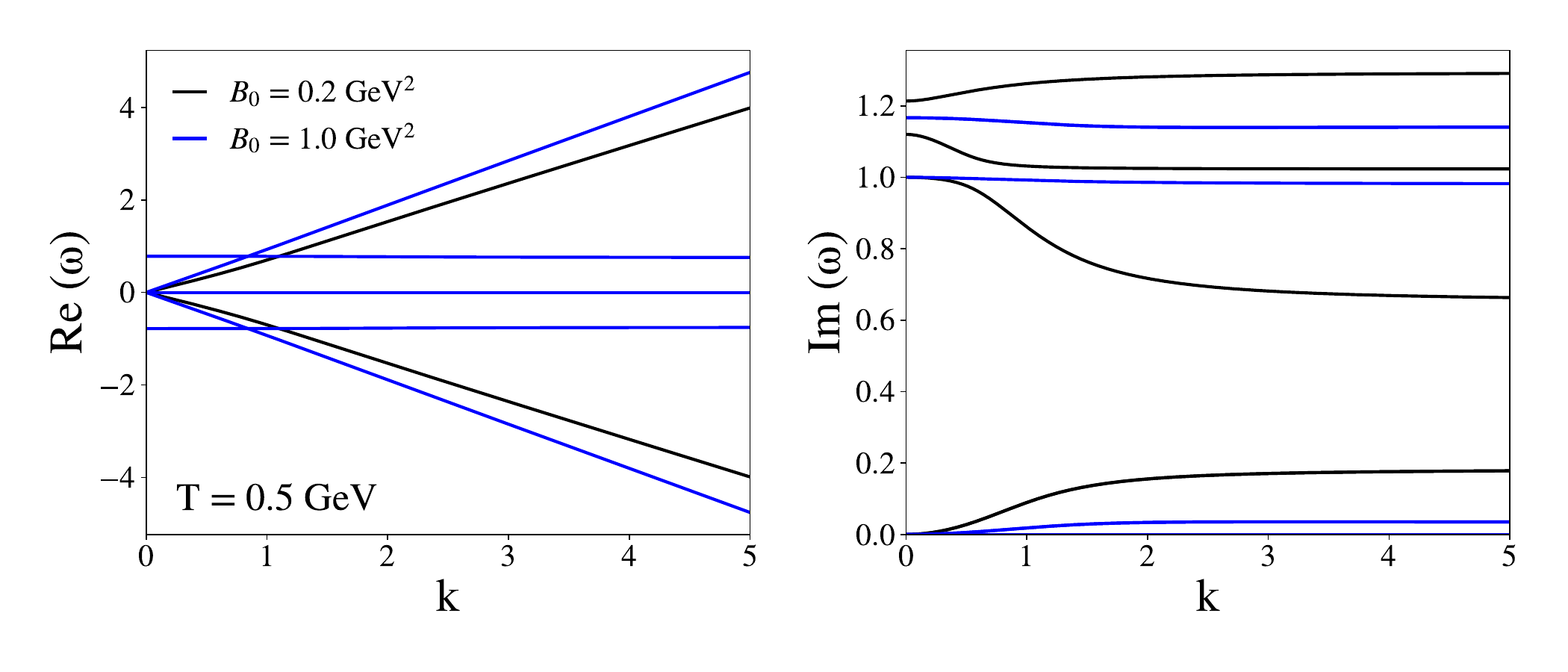} \\
\includegraphics[width=0.9\linewidth]{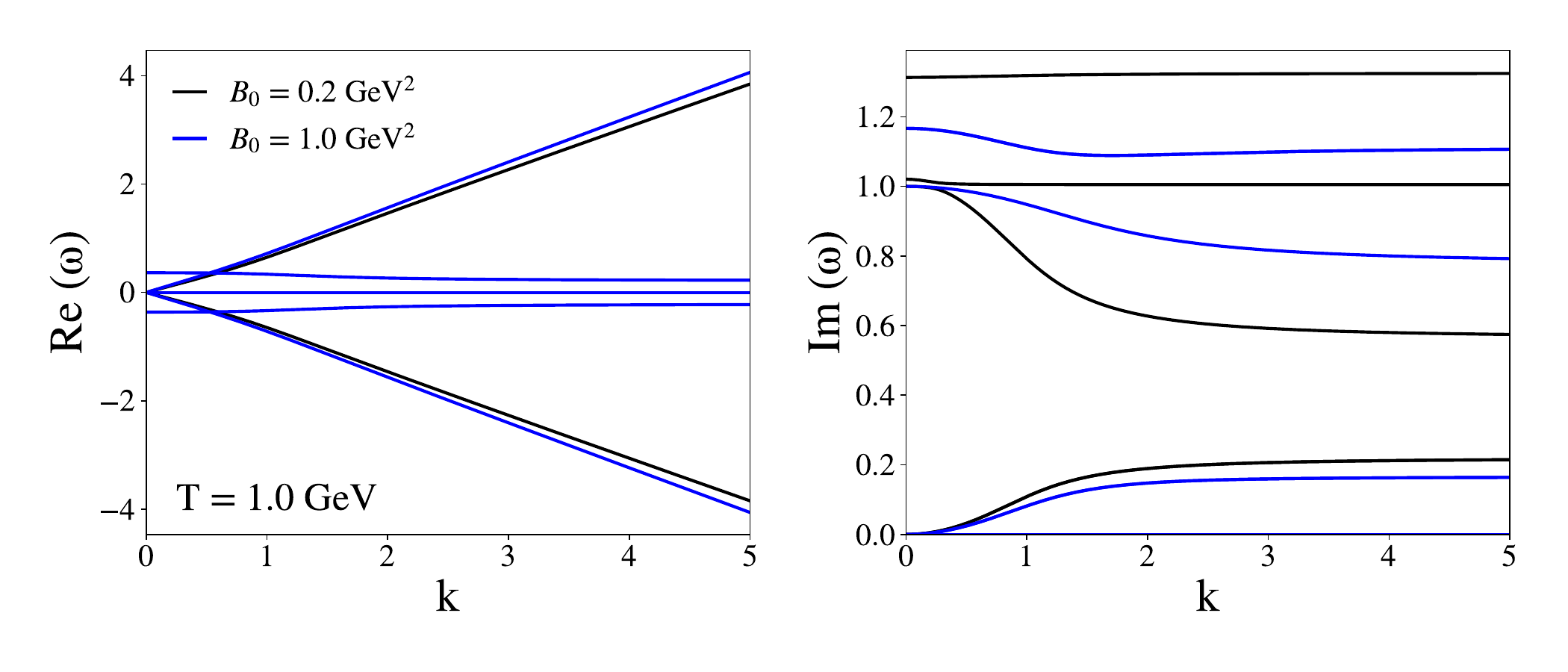}
\caption{Solutions of Eq.~\eqref{eq:trans-disp-rel-1} for $\Sigma = 1$, $\Sigma' = 4 \Sigma/3$ considering $B_0 = 0.2$ GeV$^2$ (black lines) and $B_0 = 1$ GeV$^2$ (blue lines) and $T = 0.5$ GeV (upper panels) and $T = 1$ GeV (lower panels).}
\label{fig:trans-modes-1}
\end{figure}

Finally, in the large wavenumber limit, these modes become
\begin{subequations}
\begin{align}
\omega &= \pm \frac{1}{\sqrt{5}} k + i \frac{\Sigma}{2} + \mathcal{O} \left( \frac{1}{k} \right), \label{eq:trans-largek-sound-const} \\
\omega &= \pm \sqrt{\frac{3 + 5 \mathcal{B}}{5 (1 + \mathcal{B})}} k + \mathcal{O} \left( \frac{1}{k} \right), \label{eq:trans-largek-sound-b} \\
\omega &= \pm \frac{1}{\sqrt{5 ( 1 + \mathcal{B})}} k + \frac{i \Sigma'}{2} + \mathcal{O} \left( \frac{1}{k} \right), \label{eq:trans-largek-supp-sound} \\
\omega &= i \Sigma' + \mathcal{O} \left( \frac{1}{k^2} \right), \label{eq:trans-largek-relative} \\
\omega &= i \Sigma + \mathcal{O} \left( \frac{1}{k^2} \right). \label{eq:trans-largek-total}
\end{align}
\end{subequations}
We remark that three modes have rather intricate analytical form and thus were omitted here. Nevertheless, they are displayed in Fig.~\ref{fig:trans-modes-1}.
The mode \eqref{eq:trans-largek-sound-const} propagates with a constant velocity $1/\sqrt{5}$, whereas the modes \eqref{eq:trans-largek-sound-b} and \eqref{eq:trans-largek-supp-sound} propagate with velocities that strongly depend on the magnetic field. In particular, in the limit of large magnetic fields, the first propagates with a velocity $\sim 1$, whereas the former is suppressed and becomes purely damping. Furthermore, the modes \eqref{eq:trans-largek-relative} and \eqref{eq:trans-largek-total} are both damping with timescales associated with $1/\Sigma$  and $1/\Sigma'$, respectively. The same behavior is observed in the damping sector of modes \eqref{eq:trans-largek-supp-sound} and \eqref{eq:trans-largek-sound-const}, respectively. Overall, these modes obey the linear causality condition \eqref{eq:causality-condition}, i.e., their asymptotic group velocity remains smaller than the speed of light. Since the quantities $\Sigma$ and $\Sigma'$ are positive definite, all modes are also stable for all values of magnetic field.

In Fig.~\ref{fig:trans-modes-1}, we display the solutions of Eq.~\eqref{eq:trans-disp-rel-1} for $\Sigma = 1$, $\Sigma' = 4 \Sigma/3$. We consider two different values for the magnetic field as well as for the temperature, namely $B_0 = 0.2$ GeV$^2$ (black lines) and $B_0 = 1$ GeV$^2$ (blue lines), $T = 0.5$ GeV (upper panels) and $T = 1$ GeV (lower panels). We observe that changing the temperature does not lead to qualitative changes in the behavior of these modes. Furthermore, as the magnetic field is increased, the dissipative part of the modes display a weaker dependence on the wavenumber. Last, we note that the group velocity increases with the magnetic field.

\begin{figure}[ht]%[!ht]
\centering
\includegraphics[width=0.9\linewidth]{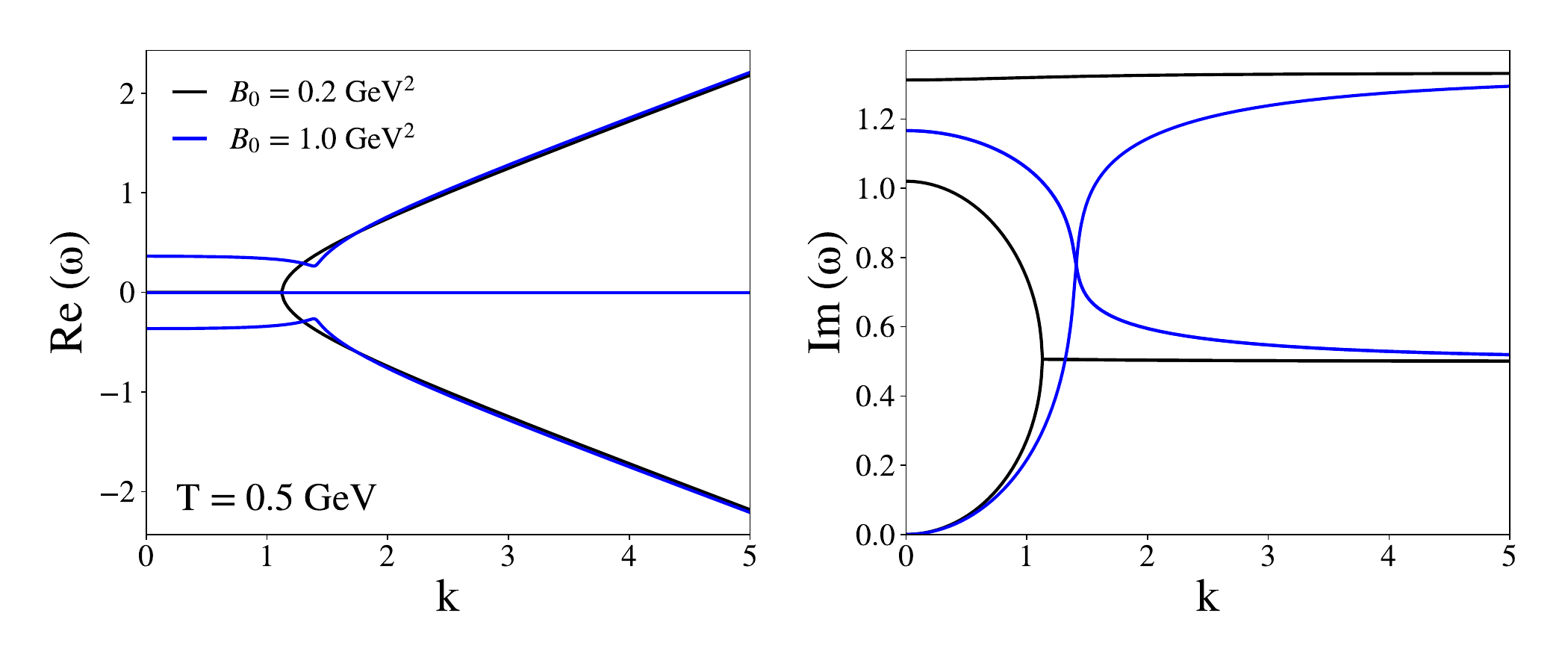} \\
\includegraphics[width=0.9\linewidth]{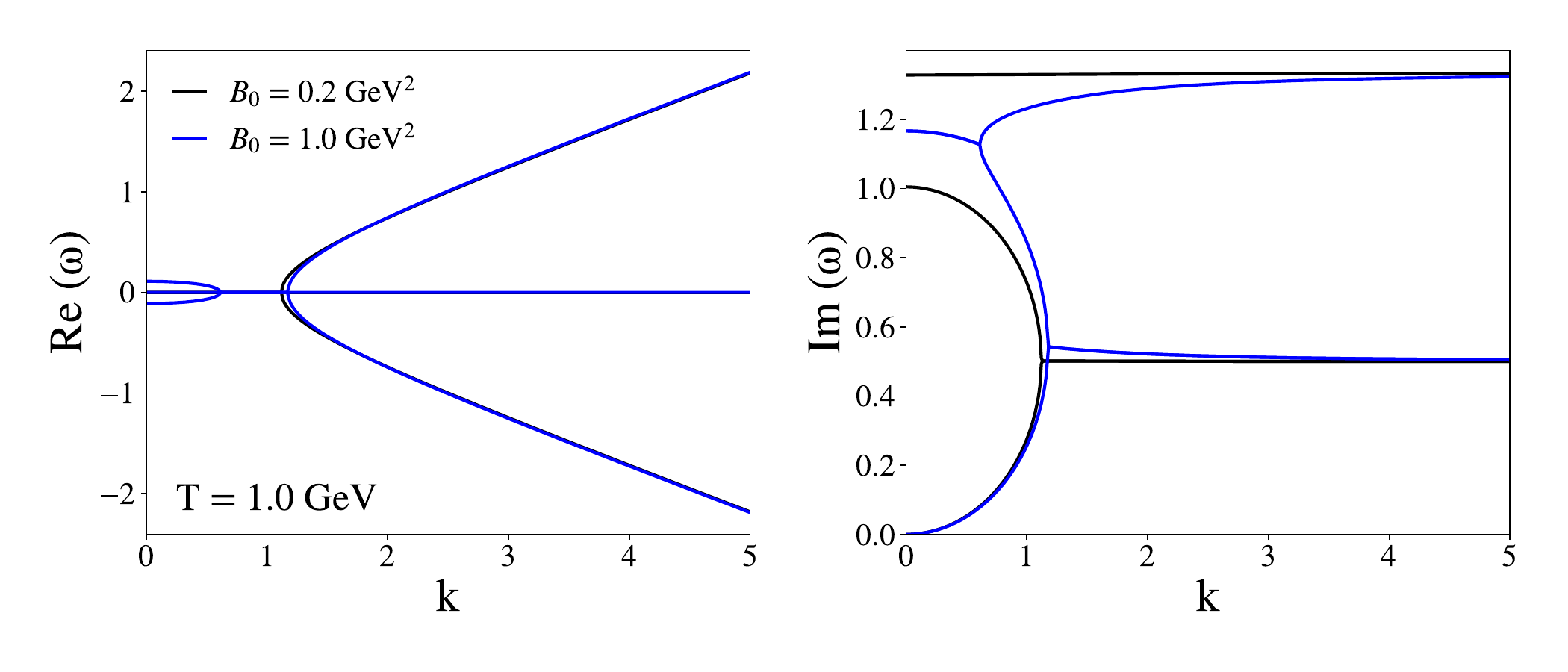}
\caption{Solutions of Eq.~\eqref{eq:trans-disp-rel-2} for $\Sigma = 1$, $\Sigma' = 4 \Sigma/3$ considering $B_0 = 0.2$ GeV$^2$ (black lines) and $B_0 = 1$ GeV$^2$ (blue lines) and $T = 0.5$ GeV (upper panels) and $T = 1$ GeV (lower panels).}
\label{fig:trans-modes-2}
\end{figure}

In Fig.~\ref{fig:trans-modes-2}, we display the solutions of Eq.~\eqref{eq:trans-disp-rel-2} for $\Sigma = 1$, $\Sigma' = 4 \Sigma/3$. Once again, we consider two values of magnetic field, $B_0 = 0.2$ GeV$^2$ (black lines) and $B_0 = 1$ GeV$^2$ (blue lines) and two values of temperature, $T = 0.5$ GeV (upper panels) and $T = 1$ GeV (lower panels). In this case, we observe that the asymptotic group velocity is rather insensitive to both the magnetic field and temperature. On the other hand, the nonhydrodynamic dissipative terms are degenerate and decreasing with the magnetic field at small values of wavenumber. However, for sufficiently large values of wavenumber, the dissipative terms tend to become identical regardless of the value of magnetic field.

\begin{figure}[ht]%[!ht]
\centering
\includegraphics[width=0.9\linewidth]{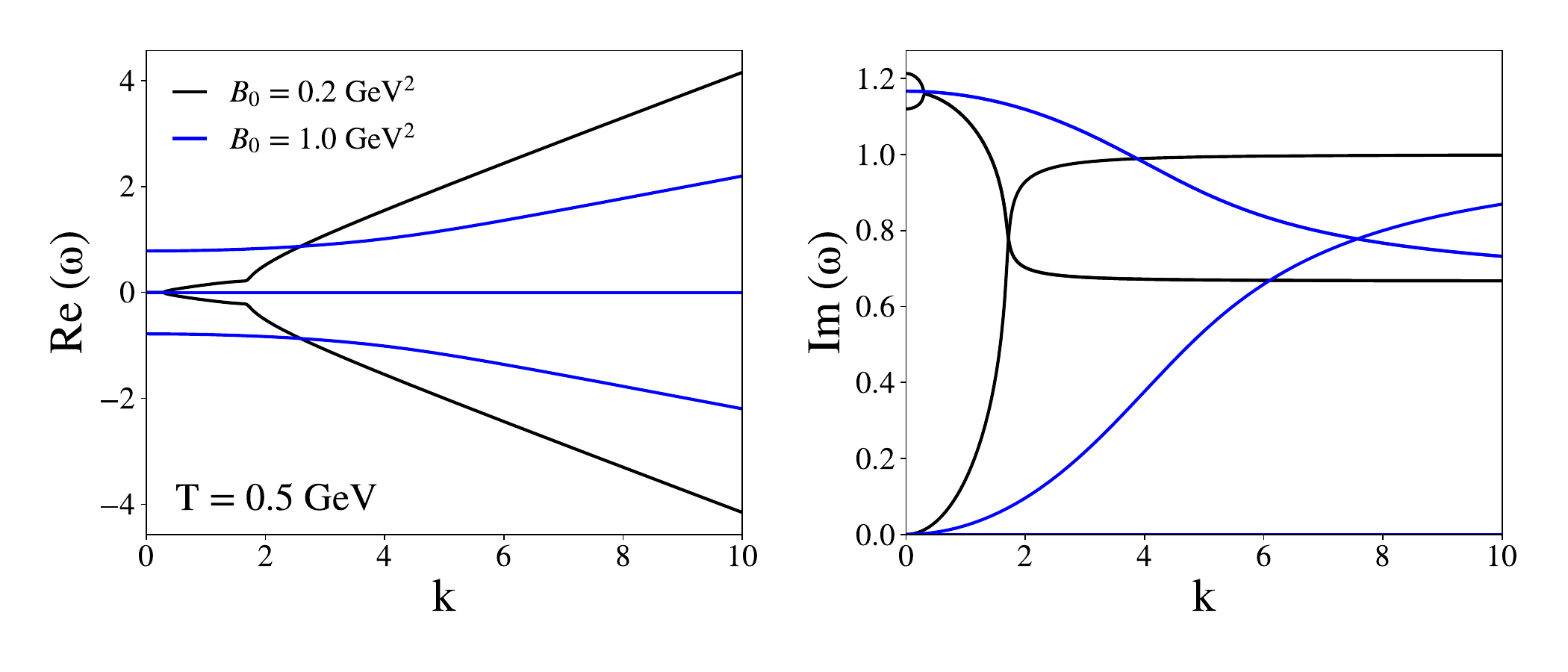} \\
\includegraphics[width=0.9\linewidth]{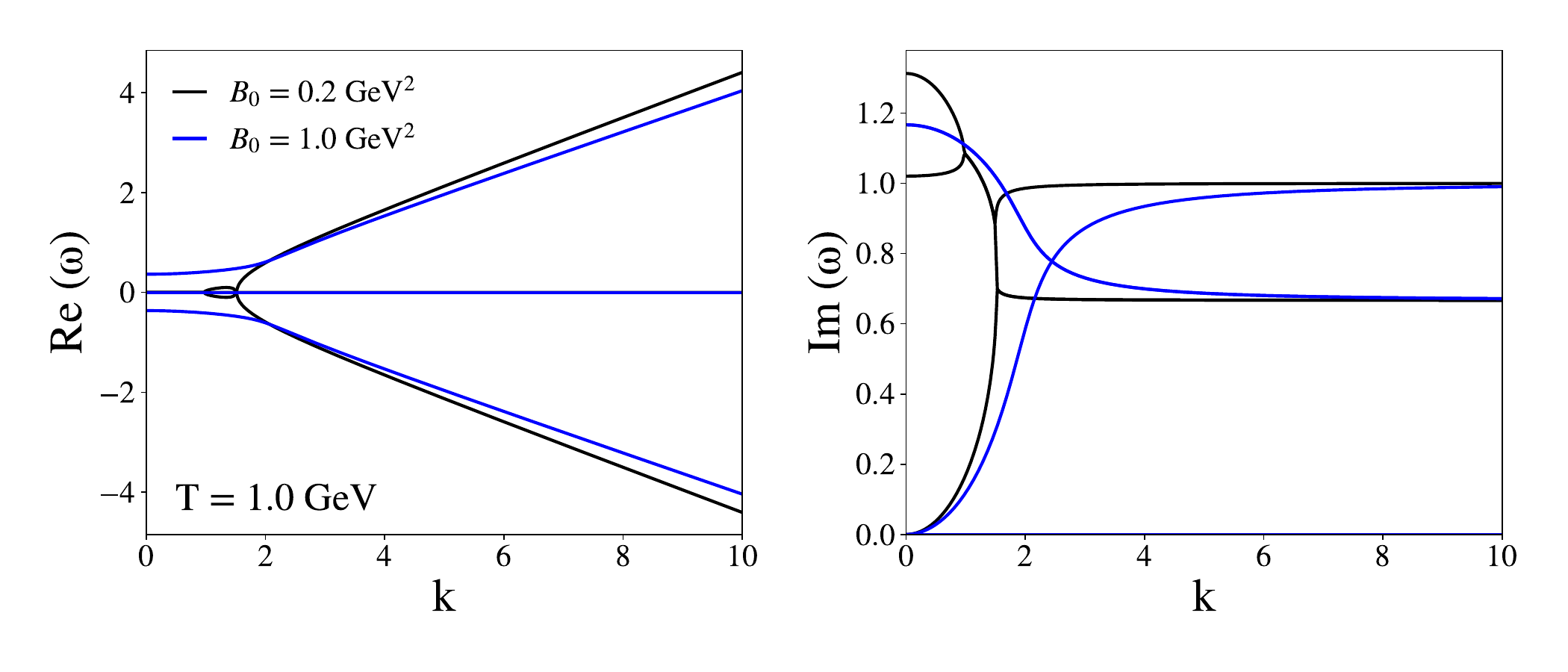}
\caption{Solutions of Eq.~\eqref{eq:pert-ub-plane-disprel-bk} for $\Sigma = 1$, $\Sigma' = 4 \Sigma/3$ considering $B_0 = 0.2$ GeV$^2$ (black lines) and $B_0 = 1$ GeV$^2$ (blue lines) and $T = 0.5$ GeV (upper panels) and $T = 1$ GeV (lower panels).}
\label{fig:trans-modes-3}
\end{figure}

In Fig.~\ref{fig:trans-modes-3}, we display the solutions of Eq.~\eqref{eq:pert-ub-plane-disprel-bk} for $\Sigma = 1$, $\Sigma' = 4 \Sigma/3$, considering $B_0 = 0.2$ GeV$^2$ (black lines) and $B_0 = 1$ GeV$^2$ (blue lines) and $T = 0.5$ GeV (upper panels) and $T = 1$ GeV (lower panels). In this case, the asymptotic group velocity decreases with increasing magnetic field. Furthermore, similarly to what as observed in Fig.~\ref{fig:trans-modes-2}, the imaginary part of the nonhydrodynamic modes are initially degenerate, but as $k$ becomes sufficiently large, such terms tend to the same value, which does not depend on the magnetic field.

We conclude that the theory is also linearly causal and stable for perturbations orthogonal to the magnetic field. Analogously to the longitudinal modes, the transverse modes always satisfy the aforementioned properties in the linear regime, regardless of the values of the transport coefficients and/or magnetic field being considered. 

\section{Longitudinal approximation of magnetohydrodynamics}
\label{sec:long-limit-mhd}

In this section, we investigate a simplified version of the magnetohydrodynamic equations, expected to be applicable in the limit of asymptotically strong magnetic fields (relative to $T^2$). When the magnetic field is large, the mean free path of particles in the direction of the magnetic field becomes larger than the Larmor radius of the particles, which determine the mean free path in the transverse directions. As argued in Ref.~\cite{Chandra:2015iza}, under these conditions, heat and momentum transport are expected to occur primarily along the magnetic field direction and the longitudinal component of the shear-stress tensor with respect to the magnetic field was considered to be the dominant contribution, allowing the transverse components to be neglected. This theory then reduces to a form similar to the Israel-Stewart framework and was recently used to study the dynamics of accretion disks around black holes, such as SgrA* and M87 \cite{Chandra:2015iza, Foucart_2015, Cordeiro:2023ljz}. 

Thus, in this regime the shear-stress tensor is assumed to only have longitudinal components with respect to the magnetic field \cite{Chandra:2015iza, Cordeiro:2023ljz, Foucart_2015},
\begin{equation}
\label{eq:trad-MHD-shear}
    \pi^{\mu \nu} \approx \pi_{bb} \left(b^\mu b^\nu + \frac{1}{2} \Xi^{\mu\nu} \right),
\end{equation}
with $\Xi^{\mu\nu} = g^{\mu\nu} - u^\mu u^\nu + b^\mu b^\nu$ being the projection operator onto the 2-space orthogonal to both $u_\mu$ and $b_\mu$. The dynamical equation satisfied by the longitudinal dissipative component $\pi_{bb}$ is determined by projecting the equations of motion \eqref{eq:coupled-PDEs-shear} with $b_\mu b_\nu$,
\begin{equation}
\label{eq:pibb}
D\pi_{bb} + \Sigma \pi_{bb} = \frac{8}{15} \varepsilon b_\mu b_\nu\sigma^{\mu \nu} + \ldots,
\end{equation}
where the ellipsis denote possible non-linear terms that were specified in Ref.~\cite{Kushwah:2024zgd}, but will not contribute to this linear analysis. We see that, for this projection of Eqs.~\eqref{eq:coupled-PDEs-shear}, the terms that depends on the magnetic field vanish and the total shear-stress tensor no longer couples to the relative shear-stress tensor, which can be simply ignored in this limit. As already mentioned, such a formalism is expected to provide a reasonable description of a plasma in the regime of strong magnetic fields \cite{Chandra:2015iza, Cordeiro:2023ljz, Foucart_2015}, where the mean free path of the particles can become significantly smaller in the directions transverse to the magnetic field.

In practice, the linearized version of these simplified equations of motion can be obtained by taking the equations derived in the previous sections and simply removing the equations of motion for the transverse components of the linearized shear-stress tensor. As mentioned, since the total shear-stress tensor does not couple to the relative shear-stress tensor, the latter can be simply ignored. Then, the relevant equations become Eqs.~\eqref{eq:cons-energ-final-rescaled}--\eqref{eq:cons-momentum-k-rescaled} and \eqref{eq:cons-mom-q-final}. 
They reduce to the following simple form,
\begin{subequations}
\label{eq:trad-mhd-limit}
\begin{align}
\Omega \Delta \hat{\varepsilon} - \kappa_{b} \Delta \tilde{u}_b - \kappa_\bot \Delta \tilde{u}_k &= 0,  \\
- \Omega \Delta u_{b} + \frac{\kappa_{b}}{3} \Delta \hat{\varepsilon} +\kappa_b \Delta \hat{\pi}_{bb} &= 0, \label{eq:trad-MHD-ub} \\
\left[ - (1 + \mathcal{B}) \Omega^2 + \mathcal{B} \left( \kappa^2_b + \kappa^2_\bot \right) \right] \Delta \tilde{u}_k + \Omega \frac{\kappa_\bot}{3} \Delta \hat{\varepsilon} &= 0, \label{eq:trad-MHD-uk} \\
\left( \Omega^{2} - \frac{\mathcal{B}}{1 + \mathcal{B}} \kappa^2_b \right) \Delta \tilde{u}_q &= 0, \label{eq:trad-q-components} \\
\left( i \Omega +\Sigma_0 \right) \Delta \hat{\pi}_{bb} - \frac{2i}{5} \left( \frac{2}{3} \kappa_b \Delta \tilde{u}_b -\frac{1}{3} \kappa_\bot \Delta \tilde{u}_k \right) &= 0. \label{eq:trad-MHD-pibb}
\end{align}
\end{subequations}
The first 4 equations above correspond to the linearized conservation laws projected into our orthonormal basis of 4-vectors, while the last equation is the linearized version of Eq.~\eqref{eq:pibb}. In the following, we analyze the modes of this simplified version of magnetohydrodynamics, considering the same cases investigated throughout the last section.

\subsection{Longitudinal perturbations}

We first consider perturbations in which $\kappa_\bot = 0$.  As before, we consider that the unperturbed plasma is at rest, which implies that $\Omega = \omega$ and $\kappa_b = k$.
In this case, Eqs.~\eqref{eq:trad-MHD-uk} and \eqref{eq:trad-q-components} yield degenerate Alfvén modes without any dissipative contribution,
\begin{equation}
\omega = \pm v_A k.
\end{equation}
That is, in this approximation of magnetohydrodynamics, these are purely propagating modes. The remaining equations lead to the following dispersion relation
\begin{equation}
\omega^3 - i \Sigma \omega^2 - \frac{3}{5} k^2 \omega + \frac{i}{3} k^2 \Sigma = 0.
\end{equation}
This dispersion relation is identical to Eq.~\eqref{eq:long-B-independent-modes} and leads to modes in the sound channel, as was already investigated and discussed in Sec.~\ref{subsec:long-pert-full-theory}. Thus, this formulation of magnetohydrodynamics is able to capture exactly one of the dispersion relations appearing in the more general description discussed in the previous section. On the other hand, it approximates the Alfv\'en modes as being purely ideal, essentially neglecting its diffusive-like contribution, see Eq.~\eqref{eq:alfvenmode}. Since such a dissipative contribution was shown to be suppressed when the magnetic field is large, this may be a reasonable approximation at least in the linear regime, when $B_0 \gg T_0^2$.  

\begin{figure}[ht]%[!ht]
\centering
\includegraphics[width=0.5\linewidth]{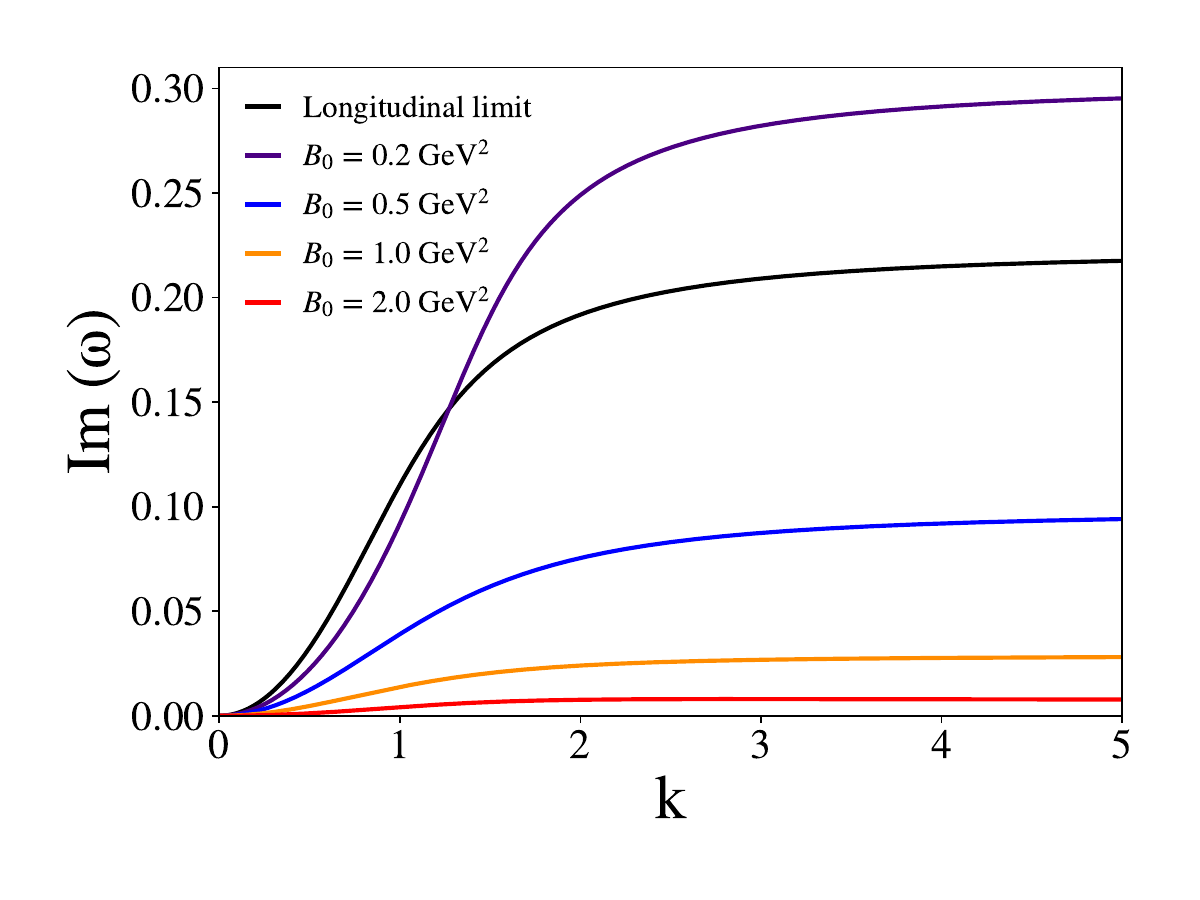}
\caption{Imaginary part of the hydrodynamic modes for longitudinal perturbations for the theory developed in Ref.~\cite{Kushwah:2024zgd} (colored lines) and for its longitudinal approximation (black line) for $B_0 = 0.2, 0.5, 1.0, 2.0$ GeV$^2$ and $T = 0.5$ GeV.}
\label{fig:comparison-long-hydro-modes}
\end{figure}

In Fig.~\ref{fig:comparison-long-hydro-modes}, we test this approximation and compare the imaginary part of the hydrodynamic modes for longitudinal perturbations, for several values of magnetic field. The black stars denote the mode that does not depend on the magnetic field and that appears in both the general theory and its simplified version containing only the longitudinal components of the shear-stress tensor. The points denote the imaginary part of the Alfv\'en mode \eqref{eq:alfvenmode} for several values of magnetic field and $T_0 = 0.5$ GeV.
We observe that the Alfv\'en modes become the dominant source of dissipation for small values of magnetic field, but are gradually suppressed as the magnetic increases. Disregarding the dynamics of the transverse components of the shear-stress tensor (with respect to the magnetic field) is only a good approximation if the magnetic field is sufficiently large to render the dissipative contribution of the Alfv\'en mode parametrically smaller than the dissipative contribution in the sound channel. We see from the figure that this starts to occur, at least for the value of temperature considered, when $B_0 \sim 2$ GeV$^2$. At the early stages of peripheral heavy ion collisions $B_0/T_0^2 \sim 0.2$ -- $2$ \cite{Hattori_2022} and, consequently, the transverse components of the shear-stress tensor should not be ignored. In accretion disks, the situation can be rather different due to the smaller temperatures, as argued in Ref.~\cite{Chen:2019usj}.

Naturally, several modes that were discussed in the previous section did not appear in this analyses, since such modes are related to the dynamics of components of the shear-stress tensor that are disregarded in the longitudinal formulation of magnetohydrodynamics. 

\subsection{Transverse perturbations}

We now consider perturbations in which $\kappa_b = 0$.  As before, we consider that the unperturbed plasma is at rest, which implies that $\Omega = \omega$ and $\kappa_\bot = k$. In this case, Eqs.~\eqref{eq:trad-MHD-ub} and \eqref{eq:trad-q-components} yield three degenerate trivial modes. The remaining equation in \eqref{eq:trad-mhd-limit} lead to the following dispersion relation,
\begin{equation}
\omega^2 \left[ \omega^3 - i \Sigma \omega^2 - \frac{1 + 3 \mathcal{B}}{3 (1 + \mathcal{B})} k^2 \omega + i \frac{1 + 3 \mathcal{B}}{3 (1 + \mathcal{B})} \Sigma k^2 \right] = 0,
\end{equation}
where two trivial modes can be identified. The non-trivial modes, on the other hand, are exactly given by
\begin{equation}
\omega = i \Sigma, \quad
\omega = \pm \sqrt{\frac{1 + 3\mathcal{B}}{3 (1 + \mathcal{B})}} k.
\end{equation}
We see that the transverse perturbations behave very different when compared to what is observed for the general theory discussed in the previous section, with the hydrodynamic modes not displaying any sign of dissipation. As discussed in the previous subsection, this may be a good approximation when the magnetic field becomes much larger than the temperature squared. 

\section{Conclusions}
\label{sec:conclusions}

In this work, we have conducted a linear stability and causality analysis around a global equilibrium state of the second-order theory of magnetohydrodynamics derived from kinetic theory \cite{Kushwah:2024zgd}, describing a binary mixture of classical, massless particles with opposite charges. The corresponding magnetohydrodynamic equations were linearized around global equilibrium and expressed in Fourier space. Next, these equations were decomposed using an orthonormal basis of 4-vectors, following the procedure developed in Ref.~\cite{Brito:2020nou}. We then studied the modes of the theory considering longitudinal and transverse perturbations (with respect to the direction of the magnetic field) on a fluid at rest. Since general solutions for the modes can be rather intricate, we restrained our analyses to the asymptotic limits of small and large values of wave number.

The formulation of magnetohydrodynamics developed in Ref.~\cite{Kushwah:2024zgd} was shown to be always causal and stable in the linear regime. Therefore, small perturbations around a locally neutral global equilibrium state described by such formulation yields modes that decrease exponentially with time and propagate subluminally. As previously mentioned, all modes have been investigated in the asymptotic limits, but also plotted for arbitrary values of wave number. We observed the occurrence of modes that are considerably distinct from those associated with traditional Israel-Stewart-like theories: at large magnetic fields, otherwise purely damped modes possess an oscillatory behavior as well. Furthermore, the Alfv\'en modes -- sound modes that propagate with the Alfv\'en velocity, $v_A$, see Eq.~\eqref{eq:alfvenmode} -- have a dissipative term that is suppressed as the magnetic field is increased. The same occurs with the other modes that arise due to perturbations that are coupled to the transverse components of the shear-stress tensor.   

We then analyzed the simplified limit of magnetohydrodynamics where only longitudinal components (with respect to $b_\mu$) of the shear-stress tensor are retained, whereas all other components are identically set to zero. This approximation is motivated by the dynamics of plasmas in strong magnetic fields, where transverse components are traditionally expected to be suppressed \cite{Chandra:2015iza}. The resulting equations decouple the shear tensor from relative shear-stress tensor and can be expressed as an Israel-Stewart-like theory.  For the longitudinal perturbations, this simplified version of the theory yields a subset of modes of those obtained for the complete theory. Additionally, it has one ideal Alfv\'en mode, which does not capture the aforementioned diffusive part of the Alfv\'en modes. Nevertheless, we have shown that the damping of such modes is suppressed when the magnetic field is sufficiently large. Furthermore, for transverse perturbations, the simplified theory yields a non-diffusive transient mode, as well as ideal sound waves, which significantly differs from the full model.

Overall, we showed that the magneto-fluid-dynamical theory derived in Ref.~\cite{Kushwah:2024csd} satisfies both causality and stability in the linear regime. In particular, these properties remain being fulfilled when the longitudinal approximation is taken. Nevertheless, we remark that such an approximation only describes part of the dynamics of the complete theory, but it does not capture dissipative behavior in both longitudinal and transverse sectors when the magnetic field is not sufficiently large. This behavior hints that the longitudinal approximation provides a good description of plasmas in the presence of strong magnetic fields, although failing to accurately capture its dynamics if the magnetic fields are not sufficiently large.

\section*{Acknowledgments}

The authors thank Jorge Noronha and Masoud Shokri for helpful discussions. C.V.P.B.~is funded by Conselho Nacional de Desenvolvimento Científico e Tecnológico (CNPq), Grant No.~140453/2021-0. K.K.~is funded by CNPq, Grant No.~163888/2021-3. G.S.D.~also acknowledges CNPq as well as Fundação Carlos Chagas Filho de Amparo à Pesquisa do Estado do Rio de Janeiro (FAPERJ), Grant No.~E-26/202.747/2018.

%%%%%%%%%%%%%%%%%%%%%%%%%%%%%%%%%%%%%%%%%%%%%%%%
%%%%%%%%%%%%%%%%%%%%%%%%%%%%%%%%%%%%%%%%%%%%%%%%
%%%%%%%%%%%%%%%%%%%%%%%%%%%%%%%%%%%%%%%%%%%%%%%%

\bibliographystyle{apsrev4-1}
\bibliography{refs}

\end{document}